\documentclass[]{aa}  

\usepackage{graphicx}
\usepackage{txfonts}
\usepackage{upgreek}
\usepackage[]{hyperref}
\usepackage{array}
\usepackage[letterspace=150]{microtype} 
\usepackage{amsmath}	
\usepackage{amssymb}	
\usepackage{xcolor}  

\begin{document} 

\title{Stellar populations across galaxy bars in the MUSE TIMER project}
\titlerunning{Stellar populations in bars}

\author{
	Justus Neumann\inst{\ref{inst1},\ref{inst2}}\thanks{E-mail: jusneuma.astro@gmail.com}
	\and Francesca Fragkoudi\inst{\ref{inst3}}
	\and Isabel P\'erez\inst{\ref{inst4},\ref{inst5}}
	\and Dimitri A. Gadotti\inst{\ref{inst6}}
	\and Jes\'us Falc\'on-Barroso\inst{\ref{inst7},\ref{inst8}}
	\and Patricia S\'anchez-Bl\'azquez\inst{\ref{inst9},\ref{inst9b}}
	\and Adrian Bittner\inst{\ref{inst6},\ref{inst10}}
	\and Bernd Husemann\inst{\ref{inst11}}
	\and Facundo A. G\'omez\inst{\ref{inst12},\ref{inst13}}
	\and Robert J. J. Grand\inst{\ref{inst3}}
	\and Charlotte E. Donohoe-Keyes\inst{\ref{inst14}}
	\and Taehyun Kim\inst{\ref{inst15},\ref{inst16}}
	\and Adriana de Lorenzo-C\'aceres\inst{\ref{inst7},\ref{inst8}}
	\and Marie Martig\inst{\ref{inst14}}
	\and Jairo M\'endez-Abreu\inst{\ref{inst7},\ref{inst8}}
	\and R\"udiger Pakmor\inst{\ref{inst3}}
	\and Marja K. Seidel\inst{\ref{inst17}}
	\and Glenn van de Ven\inst{\ref{inst6},\ref{inst11}}
}

\institute{
	Institute of Cosmology and Gravitation, University of Portsmouth, Burnaby Road, Portsmouth PO1 3FX, UK \label{inst1}
	\and Leibniz-Institut f\"ur Astrophysik Potsdam (AIP), An der Sternwarte 16, D-14480 Potsdam, Germany \label{inst2}
	\and Max-Planck-Institut f\"ur Astrophysik, Karl-Schwarzschild-Str. 1, D-85748 Garching bei M\"unchen, Germany \label{inst3}
	\and Departamento de F\'isica Te\'orica y del Cosmos, Universidad de Granada, Facultad de Ciencias (Edificio Mecenas), E-18071, Granada, Spain \label{inst4}
	\and Instituto Universitario Carlos I de F\'isica Te\'orica y Computacional, Universidad de Granada, E-18071 Granada, Spain \label{inst5}
	\and European Southern Observatory (ESO), Karl-Schwarzschild-Str.2, 85748 Garching b. M\"unchen, Germany \label{inst6}
	\and Instituto de Astrof\'isica de Canarias, Calle V\'ia L\'actea s/n, E-38205 La Laguna, Tenerife, Spain \label{inst7}
	\and Departamento de Astrof\'isica, Universidad de La Laguna, E-38200 La Laguna, Tenerife, Spain \label{inst8}
	\and Departamento de F\'isica Te\'orica, Universidad Aut\'onoma de Madrid, E-28049 Cantoblanco, Spain \label{inst9}
	\and Instituto de F\'isica de Part\'iculas y del Cosmos IPARCOS, Facultad de Ciencias F\'isicas, Universidad Complutense de Madrid, E-28040 Madrid, Spain \label{inst9b}
	\and Ludwig-Maximilians-Universit\"at, Professor-Huber-Platz 2, 80539 M\"unchen, Germany \label{inst10}
	\and Max-Planck-Institut f\"ur Astronomie, K\"onigstuhl 17, D-69117 Heidelberg, Germany \label{inst11}
	\and Instituto de Investigaci\'on Multidisciplinar en Ciencia y Tecnolog\'ia, Universidad de La Serena, Ra\'ul Bitr\'an 1305, La Serena, Chile \label{inst12}
	\and Departamento de Astronom\'ia, Universidad de La Serena, Av. Juan Cisternas 1200 Norte, La Serena, Chile \label{inst13}
	\and Astrophysics Research Institute, Liverpool John Moores University, IC2, Brownlow Hill, Liverpool, Merseyside L3 5RF, UK \label{inst14}
	\and Department of Astronomy and Atmospheric Sciences, Kyungpook National University, Daegu 702-701, Korea \label{inst15}
	\and Korea Astronomy and Space Science Institute, Daejeon 305-348, Korea \label{inst16}
	\and Caltech-IPAC, MC 314-6, 1200 E California Blvd, Pasadena, CA, 91125, USA \label{inst17}
}

\date{Received 29/01/2020; accepted 18/03/2020}

\abstract{
Stellar populations in barred galaxies save an imprint of the influence of the bar on the host galaxy's evolution. We present a detailed analysis of star formation histories (SFHs) and chemical enrichment of stellar populations in nine nearby barred galaxies from the TIMER project. We use integral field observations with the MUSE instrument to derive unprecedented spatially resolved maps of stellar ages, metallicities, [Mg/Fe] abundances and SFHs, as well as H$\upalpha$ as a tracer of ongoing star formation. We find a characteristic V-shaped signature in the SFH perpendicular to the bar major axis which supports the scenario where intermediate age stars ($\sim 2$-$6\ \mathrm{Gyr}$) are trapped on more elongated orbits shaping a thinner part of the bar, while older stars ($> 8\ \mathrm{Gyr}$) are trapped on less elongated orbits shaping a rounder and thicker part of the bar. We compare our data to state-of-the-art cosmological magneto-hydrodynamical simulations of barred galaxies and show that such V-shaped SFHs arise naturally due to the dynamical influence of the bar on stellar populations with different ages and kinematic properties. Additionally, we find an excess of very young stars ($< 2\ \mathrm{Gyr}$) on the edges of the bars, predominantly on the leading side, confirming typical star formation patterns in bars. Furthermore, mass-weighted age and metallicity gradients are slightly shallower along the bar than in the disc likely due to orbital mixing in the bar. Finally, we find that bars are mostly more metal-rich and less [Mg/Fe]-enhanced than the surrounding discs. We interpret this as a signature that the bar quenches star formation in the inner region of discs, usually referred to as star formation deserts. We discuss these results and their implications on two different scenarios of bar formation and evolution.
}


\maketitle



%

\section{Introduction}

Most disc galaxies in the nearby Universe are barred, with numerous observational studies finding fractions on the order of 60-80\% \citep{Eskridge2000,Menendez-Delmestre2007,Aguerri2009,Masters2011,Buta2015,Erwin2018}. In principle, one would expect this number to be even larger, because bars are long-lived \citep{Gadotti2015} and it is extremely difficult to avoid bar-forming instabilities in numerical simulations of disc galaxies \citep[e.g.][]{Berrier2016,Bauer2019}. These statistics already demonstrate that studying bars is essential for the global understanding of galaxy evolution.

Bars are very efficient in the radial redistribution of matter and angular momentum and thereby drive the formation of nuclear structures such as inner bars \citep{Lorenzo-Caceres2012,Lorenzo-Caceres2013}, nuclear rings or nuclear discs \citep{Debattista2006,Athanassoula2013a,Sellwood2014,Fragkoudi2019b}, as well as outer structures like inner and outer rings \citep[see also][]{Buta1986,Buta1996,Rautiainen2000}. At the same time they are believed to intensify the global cessation of star formation in late stages of galaxy evolution \citep{Masters2012,Hakobyan2016,Haywood2016,Khoperskov2018,George2019}. They may play a role in feeding active galactic nuclei (AGN) by transporting gas inwards, but this is a heavily discussed subject and still somewhat inconclusive \citep{Ho1997,Coelho2011,Oh2012,Cheung2015,Galloway2015,Goulding2017,Alonso2018}. 

The impact of bars in shaping their host galaxies has been studied in detail in the literature. In contrast, quantitative observational studies of internal properties of bars, such as star formation and stellar populations, are still scarce. Observations of stellar populations in bars provide information about processes during bar formation and evolution. Among possible sources for a variation of stellar populations are localised star formation during certain periods in time, radial migration of stars, quenching of star formation, and other dynamics of stars and gas.

Most observational studies so far have been focused on stellar population gradients along the bar major axis as compared to the outer disc or to the minor axis of the bar \citep{Perez2007,Perez2009,Perez2011,Sanchez-Blazquez2011,Seidel2016,Fraser-McKelvie2019} or compared to an unbarred control sample \citep{Sanchez-Blazquez2014}. From theory we could expect a flattening of mean stellar parameters along the major axis. From stellar dynamics, we know that stars get trapped in periodic and quasi-periodic orbits in the bar potential. However, \citet{Sellwood2002} have shown that, when spiral arms are present in a galaxy, stars can gain or lose angular momentum at the corotation resonance without heating the disc \citep[see also][]{Grand2012,Halle2015,Halle2018}. This process can be enhanced by coupling with a bar potential \citep{Minchev2010}. As a consequence, stars migrate radially which would result in a flattening of the stellar chemical abundance gradient \citep{Grand2015}. With a growing bar this process can affect large parts of the disc, but apparently in simulations it is mostly visible outside corotation, i.e. outside the bar region \citep[e.g.][]{Friedli1994, DiMatteo2013}. Additionally, gradients along bars are expected to be flat due to orbital mixing \citep{Binney1987}. Bars are confined elongated structures. In any spatial resolution element within the bar, stellar orbits with different elongations and apocentres cross or come very close together. This would result in a mixing and a flattening of measured stellar population gradients along the bar. From observational studies, the results still seem somehow ambiguous, but they mostly indicate a flattening along the major axis \citep{Sanchez-Blazquez2011,Seidel2016,Fraser-McKelvie2019}.

However, the distribution of stellar populations is influenced by more factors. From hydrodynamical simulations of gas dynamics in barred galaxies, we know that gas flows inwards in thin stream lines along the leading edges\footnote{In this paper, we call the two long sides of a bar \emph{edges},
if a bar in 2D projection is thought of as a rectangle. The \emph{leading edges} are those that are on the forefront of the rotating bar. The two short sides, we will address as \emph{ends} of the bar.} of rotating bars \citep{Athanassoula1992b,Piner1995,KimWT2012,Li2015,Renaud2015b,Sormani2015,Fragkoudi2016}. If gas is present and star formation is not suppressed by shear, star formation is expected to happen along the leading edges of the bar. This was observationally confirmed in \citet{Neumann2019}, where we found that only some bars show signs of ongoing star formation and it is predominantly located on the leading side \citep[see also][]{Sheth2002a}. Such a pattern could be observed in the youngest stellar populations but is expected to be washed out fast due to orbital mixing and short dynamical timescales ($\sim \mathrm{100\,Myr}$).

Additionally, a local cessation of star formation would also leave its imprints on the stellar populations in the form of a truncated star formation history (SFH) or elevated [Mg/Fe] values, the latter of which is commonly used as a time-scale indicator of the SFH. \citet{Seidel2016} found that the main disc is usually less $\upalpha$-enhanced than the bar indicating a more continuous star formation, while more central parts of the disc have been observed with truncated SFHs due to the action of bars \citep{James2016, James2018}.

Finally, recent $N$-body simulations have shown that stars could be trapped into bar orbits with different elongations based on the initial kinematics of the stars or the gas out of which they form \citep{Athanassoula2017,Debattista2017,Fragkoudi2017}. This could lead to different populations dominating at different locations in the bar.

In this work, we present spatially resolved stellar populations analyses of nine barred galaxies from the {\bf T}ime {\bf I}nference with {\bf M}USE in {\bf E}xtragalactic {\bf R}ings (TIMER) project studied with the Multi-Unit Spectroscopic Explorer \citep[MUSE;][]{Bacon2010} on the Very Large Telescope (VLT). We specifically concentrate on the bar region, while other components of the galaxies will be analysed in future papers by the collaboration. Additionally to stellar ages, metallicities and [Mg/Fe] abundances, we present a detailed analysis of SFHs across bars and we use H$\upalpha$ measurements to connect young stellar populations to places of ongoing star formation. Furthermore, we compare our results to the bars in magneto-hydrodynamical cosmological simulations of the Auriga project \citep{Grand2017}.

With the high spatial resolution of our data, we are able to resolve the bar not only along the major axis but also across its width. In this paper, we explore the stellar populations of bars to better understand processes during their formation and evolution that include star formation, quenching, radial migration and kinematic differentiation.

The outline of the paper is as follows. In Sect. \ref{sect:data}, we present the TIMER sample, the selection of our subsample and the observations with MUSE, as well as the emission line and stellar population analysis. In Sect. \ref{sect:maps}, we show 2D maps of spatially resolved H$\upalpha$ and stellar population properties, followed by an analysis of gradients along 1D pseudo-cuts extracted from the maps in Sect. \ref{sect:gradients}. In Sect. \ref{sect:SFH}, we present results from detailed SFHs. Finally, we discuss some of the most important results in Sect. \ref{sect:discussion}, where we also compare our observations to simulations from the Auriga project, and conclude the work in Sect. \ref{sect:conclusion}.

\section{Data and analysis}
\label{sect:data}

\subsection{Sample selection and MUSE observations}
\label{sect:sample_and_obs}

\defcitealias{Gadotti2019}{Paper I}

The present work is part of the TIMER project \citep[hereafter \citetalias{Gadotti2019}]{Gadotti2019}, a survey with the MUSE integral field unit (IFU) spectrograph that aims at studying the central structures of 24 nearby barred galaxies. One of the main goals of the project is to estimate the epoch when galactic discs dynamically settle, which leads to the formation of bars. The feasibility was demonstrated in a pilot study of the galaxy NGC4371 \citep{Gadotti2015}. Within the TIMER collaboration, TIMER data has been used to study the assembly of double-barred galaxies \citep{Lorenzo-Caceres2019a}, to find that inner bars also buckle \citep{Mendez-Abreu2019} and to explore bar-driven effects on the interstellar medium, central star formation and stellar feedback \citep{Leaman2019}.

The parent sample of the TIMER project is that of the Spitzer Survey of Stellar Structures in Galaxies \citep[$\mathrm{S^{4}G}$;][]{Sheth2010} and includes only nearby ($d<40\,\mathrm{Mpc}$), bright ($m_\mathrm{B}<15.5$) and large ($D_{25}>1\arcmin$) galaxies. From this catalogue, TIMER galaxies were selected based on mass ($M_{\star}>10^{10}\,M_{\sun}$), inclination ($i<60\degr$) and the presence of a bar and prominent central structures, such as nuclear rings or nuclear discs. The latter was judged consulting the morphological classification in \citet{Buta2015}.

\begin{figure}
	\centering
	\includegraphics[width=\columnwidth]{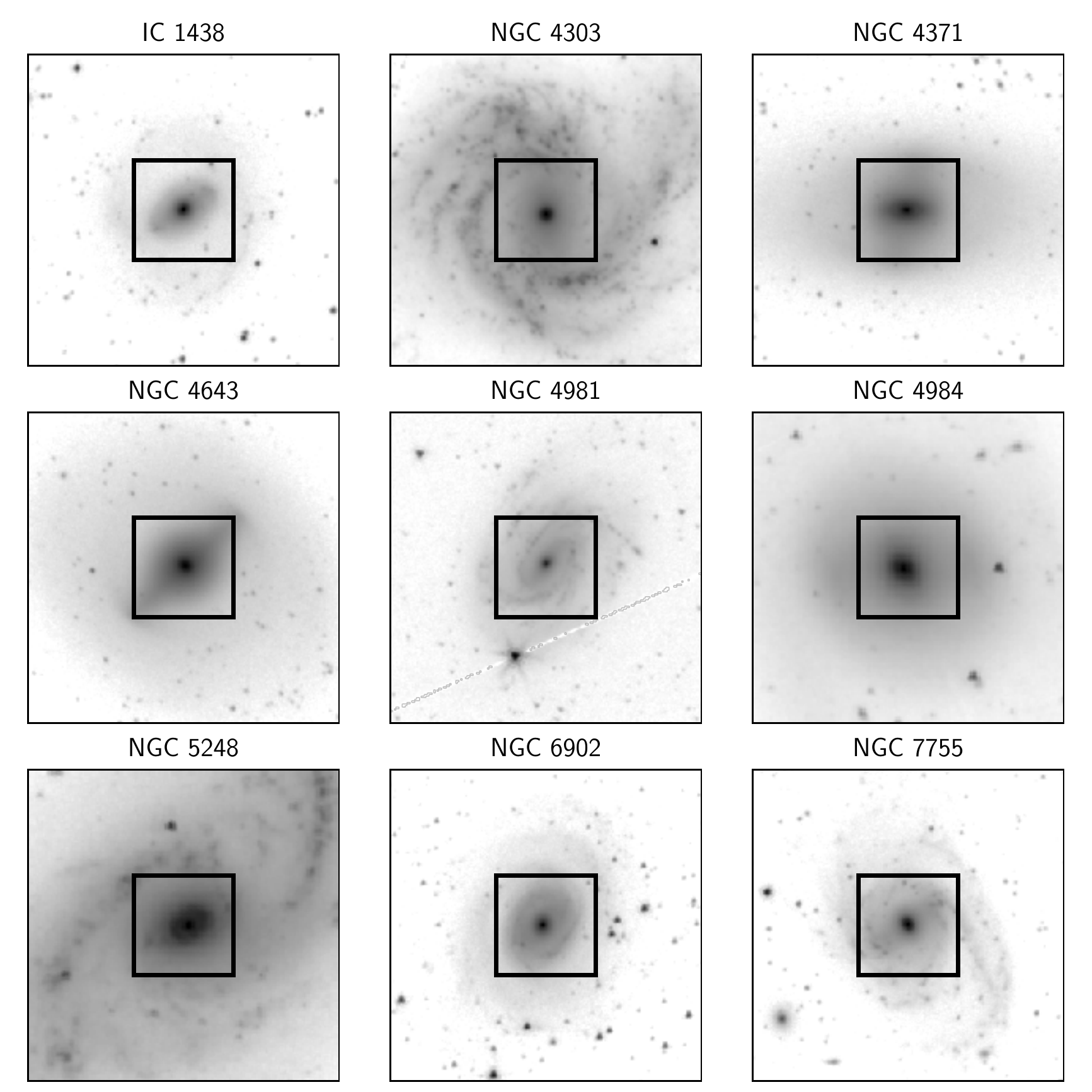}
	\caption{$\mathrm{S^{4}G}$ images at $3.6\,\mathrm{\upmu m}$ for the complete sample. The black squares show the approximate MUSE FOV.}
	\label{fig:sample_images}
\end{figure}

Out of the 24 nearby barred galaxies in the TIMER sample, 21 galaxies have been observed with MUSE up to date. From these 21 objects we selected all galaxies where almost the entire bar ($80\%$ of the semi-major axis of the bar) is covered by the MUSE field-of-view (FOV) to be able to study gradients along bars. The final sample consists of nine galaxies, the bars of six of them completely fit into the MUSE FOV. In Fig. \ref{fig:sample_images}, we show infrared $3.6\,\mathrm{\upmu m}$ $\mathrm{S^{4}G}$ images of the sample superimposed with the approximate outline of the MUSE FOV. The main parameters of the sample are summarised in Table \ref{tbl:sample}. This table includes parameters for the bars from the $\mathrm{S^{4}G}$ analyses to constrain the bar region (length, ellipticity and position angle) from \citet{Herrera-Endoqui2015}\footnote{Using $3.6\,\upmu \mathrm{m}$ images from \emph{Spitzer}, bar lengths were measured visually, while orientation and ellipticity were determined interactively by visually marking the object and fitting ellipses to the marked points. See \citet{Herrera-Endoqui2015} for more detail.} and the bar strength from \citet{Diaz-Garcia2016} which we will use to compare to stellar population parameters derived from the TIMER observations.

Observations of eight of the galaxies were performed during ESO Period 97 from April to September 2016. NGC 4371 was subject to our science verification programme for MUSE \citep{Gadotti2015} and observed between the 25th and the 29th of June 2014. The MUSE instrument covers a 1 squared arcmin FOV with a spatial sampling of $0.2\arcsec$ and a spectral sampling of $1.25\,$\AA\ per pixel. We used the nominal setup with a wavelength coverage from $4750\,$\AA\ to $9350\,$\AA\ at a mean resolution of $2.65\,$\AA\ (full-width-at-half-maximum, FWHM). The typical seeing during observations was $0.8\arcsec$-$0.9\arcsec$. The data was reduced with the MUSE pipeline v1.6 \citep{Weilbacher2012} applying the standard calibration plan. Details of the TIMER sample selection, the observations and the data reduction can be found in \citetalias{Gadotti2019}.

\begin{table*}
\caption{Summary of the main parameters of the sample.}
\label{tbl:sample}
\centering

\begin{tabular}{l l r r r r r r r r}
\hline\hline
Galaxy & Type & $i$ & $d$ & $M_\star$ & $M_\mathrm{HI}$ & $\mathrm{Q_{bar}}$ & $L_\mathrm{{bar}}$ & $PA_\mathrm{{bar}}$ & $\epsilon_\mathrm{{bar}}$ \\
 & & deg & Mpc & $10^{10}\,\mathrm{M}_\odot$ & $10^{10}\,\mathrm{M}_\odot$ & & arcsec & deg & \\
(1) & (2) & (3) & (4) & (5) & (6)& (7)& (8)& (9)& (10)\\
\hline
  IC 1438 & $\mathrm{(R_1)SAB_a(r',\underline{l},nl)0/a}$ & 24 & 33.8 & 3.1 & 0.12 & 0.178 & 23.8 & 121.0 & 0.51\\
  NGC 4303 & $\mathrm{SAB(rs,nl)b\underline{c}}$ & 34 & 16.5 & 7.2 & 0.45 & 0.535 & 36.1 & 178.0 & 0.69\\
  NGC 4371 & $\mathrm{(L)SB_a(r,bl,nr)0^{0/+}}$ & 59 & 16.8 & 3.2 & 0.08 & 0.234 & $34.8^\star$ & $159.0^\star$ & $0.51^\star$\\
  NGC 4643 & $\mathrm{(L)SB(\underline{r}s,bl,nl)0^{0/+}}$ & 44 & 25.7 & 10.7 & 0.03 & 0.272 & 49.9 & 133.0 & 0.47\\
  NGC 4981 & $\mathrm{SA\underline{B}(s,nl)\underline{b}c}$ & 54 & 24.7 & 2.8 & 0.35 & 0.093 & 18.9 & 147.0 & 0.57\\
  NGC 4984 & $\mathrm{(R'R)SAB_a(l,bl,nl)0/a}$ & 53 & 21.3 & 4.9 & 0.03 & 0.176 & 30.0 & 94.0 & 0.30\\
  NGC 5248 & $\mathrm{(R')SAB(s,nr)bc}$ & 41 & 16.9 & 4.7 & 0.40 & 0.138 & 27.4 & 128.0 & 0.36\\
  NGC 6902 & $\mathrm{(R')S\underline{A}B(\underline{r}s,nl)\underline{a}b}$ & 37 & 38.5 & 6.4 & 2.34 & 0.045 & 16.2 & 132.5 & 0.36\\
  NGC 7755 & $\mathrm{(R')SAB(rs,nrl)\underline{b}c}$ & 52 & 31.5 & 4.0 & 0.65 & 0.401 & 24.6 & 125.0 & 0.56
\\
\hline
\end{tabular}
\tablefoot{
Column (1)-(6) are extracted from Table 1 in \citetalias{Gadotti2019}. For details, we refer to that paper. (1) Galaxy name; (2) morphological type by \citet{Buta2015}; (3) inclination of the galaxy; (4) distance to the galaxy; (5) total stellar mass; (6) total \ion{H}{I} mass; (7) bar maximum gravitational torque from \citet{Diaz-Garcia2016}. Column (8)-(10) are taken from \citet{Herrera-Endoqui2015} with the exception of NGC 4371, which was found to be inaccurate and taken from \citet{Gadotti2015} instead: (8) bar length; (9) position angle of the bar; (10) ellipticity of the bar.}
\end{table*}

\subsection{Emission line analysis}
\label{sect:emission_line_analysis}

The extraction of emission line fluxes for all TIMER galaxies was performed by employing the code \textls{\sc PyParadise}, an extended version of \textls{\sc paradise} \citep[see][]{Walcher2015}. One of the advantages of \textls{\sc PyParadise} is that it propagates the error from the stellar absorption fit to the emission line analysis. The procedure was done on a spaxel-by-spaxel basis to retrieve the fine spatial structure of the gas component. This is possible owing to the generally high signal-to-noise (S/N) in the emission lines. The stellar absorption features, however, are usually less pronounced. For that reason, we Voronoi binned the cubes with minimum S/N of $\sim 40$ to estimate the underlying stellar kinematics. For self-consistency and to make use of the internal error propagation, we did not use the kinematics derived with \textls{\sc pPXF} that we describe in the next subsection, but performed an independent analysis with \textls{\sc PyParadise}.

The procedure can be summarised in three steps, further details can be found in \citetalias{Gadotti2019}. First, the stellar kinematics are measured by fitting a linear combination of stellar template spectra from the Indo-US template library \citep{Valdes2004} convolved with a Gaussian line-of-sight velocity kernel to the Voronoi-binned spectra in the cube. Second, in each spaxel, the continuum is fitted with fixed kinematics according to the underlying Voronoi cell. Finally, the emission lines are modelled with Gaussian functions in the continuum-subtracted residual spectra. To estimate uncertainties, the fit is repeated 30 times in a Monte Carlo simulation after modulating the input spectra by the formal errors and by using only 80\% of the template library.

The extracted H$\upalpha$ fluxes have to be corrected for dust attenuation. For that purpose, we used the ratio of H$\upalpha$/H$\upbeta =2.86$ (Balmer decrement from case B recombination), which is intrinsically set by quantum mechanics. Since the attenuation is wavelength dependent, the observed ratio changes and can thus be used to correct for the effect of dust on the emission line fluxes. We used the prescription by \citet{Calzetti2000} to account for the wavelength dependent reddening.

\subsection{Derivation of stellar population parameters}
\label{sect:SP_deriv}

A detailed description of the extraction of stellar population parameters for the whole set of TIMER galaxies is given in \citetalias{Gadotti2019}. Here we summarise the main steps of the procedure.

To secure a high quality of the analysis, the spectra in each cube were spatially binned using the Voronoi method of \citet{Cappellari2003} to achieve a minimum S/N of $\sim 40$ per spatial element. The spectrum of each Voronoi bin was then analysed as follows.

First, the stellar kinematics were determined employing the penalized pixel fitting code \textls{\sc pPXF} \citep{Cappellari2004,Cappellari2017} with the E-MILES single stellar population (SSP) model library from \citet{Vazdekis2015}. Subsequently, with fixed stellar kinematics, the nebular emission was fitted and removed with the code \textls{\sc gandalf} \citep{Sarzi2006, Falcon-Barroso2006}. Afterwards, we modelled ages, metallicities and SFHs on the emission-free residual spectra employing the code \textls{\sc steckmap} \citep[STEllar Content and Kinematics via Maximum A Posteriori;][]{Ocvirk2006a,Ocvirk2006b} with the E-MILES library and assuming a \citet{Kroupa2001} initial mass function (IMF). We employed the BaSTI isochrones \citep{Pietrinferni2004,Pietrinferni2006,Pietrinferni2009, Pietrinferni2013} with stellar ages ranging from $0.03$-$14.0\,$Gyr and metallicities (Z) from 0.0001 to 0.05, corresponding to [Z/H] ranging from $-2.3$ to $0.4$. We refer to \citetalias{Gadotti2019} for further technical details.

Uncertainties in the derivation of mean stellar ages and metallicities from the SFHs produced by \textls{\sc steckmap} were studied for a set of 5000 spectra from the TIMER data in Appendix A of \citetalias{Gadotti2019}. Typical values are $0.5$-$1\,$Gyr for age, and $0.005$-$0.010$ for metallicity (Z). 

Since \textls{\sc steckmap} is not capable of measuring [Mg/Fe] abundances, we exploit the \textls{\sc pPXF} routine to derive those values in a similar but slightly optimised set-up \citep[see, e.g.][]{Pinna2019a}. The implementation of this analysis is based on the \textls{\sc gist} pipeline\footnote{\url{http://ascl.net/1907.025}} \citep{Bittner2019} and further details of the analysis are described in Bittner et al. (in prep.). A comparison between the results obtained from \textls{\sc steckmap} and \textls{\sc pPXF} is currently conducted within the TIMER collaboration and will be published soon. Differences are found to be minimal. In the following, we summarise the main steps of exploiting the \textls{\sc pPXF} routine.

In order to obtain reliable estimates of the [Mg/Fe] values, in this analysis, we spatially bin the data to an approximately constant S/N of 100. We note that all spaxels which surpass this S/N remain unbinned while those below the isophote level that has an average S/N of 3 are excluded from the analysis. As line-spread function of the observations we adopt the udf-10 characterisation of \citet{Bacon2017}. 

We employ the wavelength range of $4800\,$\AA$\,$to $5800\,$\AA$\ $together with the MILES model library from \citet{Vazdekis2010}, covering a large range in ages and metallicities, and two $\mathrm{[\upalpha/Fe]}$ values of 0.00 and 0.40. In the given wavelength range, the best-fit combination of templates with regard to their [$\upalpha$/Fe] value is driven only by the Mg lines and, since it is not clear if all $\upalpha$-elements are enhanced at the same level, we will, thus, refer to this abundance in the following as [Mg/Fe]. The models employ the BaSTI isochrones \citep{Pietrinferni2004,Pietrinferni2006,Pietrinferni2009,Pietrinferni2013} and the revised Kroupa initial mass function \citep{Kroupa2001}. In order to account for differences between observed and template spectra, we include an 8th-order, multiplicative Legendre polynomial. 

The analysis is performed in three steps: We first derive the stellar kinematics with \textls{\sc pPXF} with emission lines masked, before modelling and subtracting any gaseous emission with \textls{\sc pyGandALF} -- a python implementation of \textls{\sc gandalf}. Then, we perform a regularised run of \textls{\sc pPXF} to estimate the population properties, while keeping the stellar kinematics fixed to the results from the unregularised run.  The used strength of the regularisation is the one at which the $\chi^2$ of the best-fitting solution of the regularised run exceeds the one from the unregularised run by $\sqrt{2N_{\mathrm{pix}}}$, with $N_{\mathrm{pix}}$ being the number of spectral pixels included in the fit. This criterion is applied to one of the central bins with high S/N and then used for the entire cube \citep{Press1992,McDermid2015}. 

\section{Resolved 2D properties}
\label{sect:maps}

\subsection{Recent star formation as traced by H$\alpha$}
\label{sect:Halpha}

To get a complete picture of the stellar population properties in bars it is important to connect the study of the SFH with ongoing star formation. A detailed investigation of star formation in bars was conducted for a different sample in \citet{Neumann2019}. There, we found that bars clearly separate into either star-forming or non-star-forming and, if star formation is present, it is predominant on the leading edge of the rotating bar. In the present work, we explore how star formation is connected to the stellar populations in the bar.

From the MUSE data cubes, we derived H$\upalpha$ maps as tracer of \ion{H}{II} regions and, thus, star formation for the complete set of galaxies. Given that gas can also be ionised by AGN or shocks, we derived Baldwin, Phillips, \& Terlevich diagrams  \citep[BPT\ diagrams;][]{Baldwin1981} that showed that H$\upalpha$ emission in the bar and the disc is not affected by the AGN and can safely be accounted to star formation.

\begin{figure}
	\centering
	\includegraphics[width=\columnwidth]{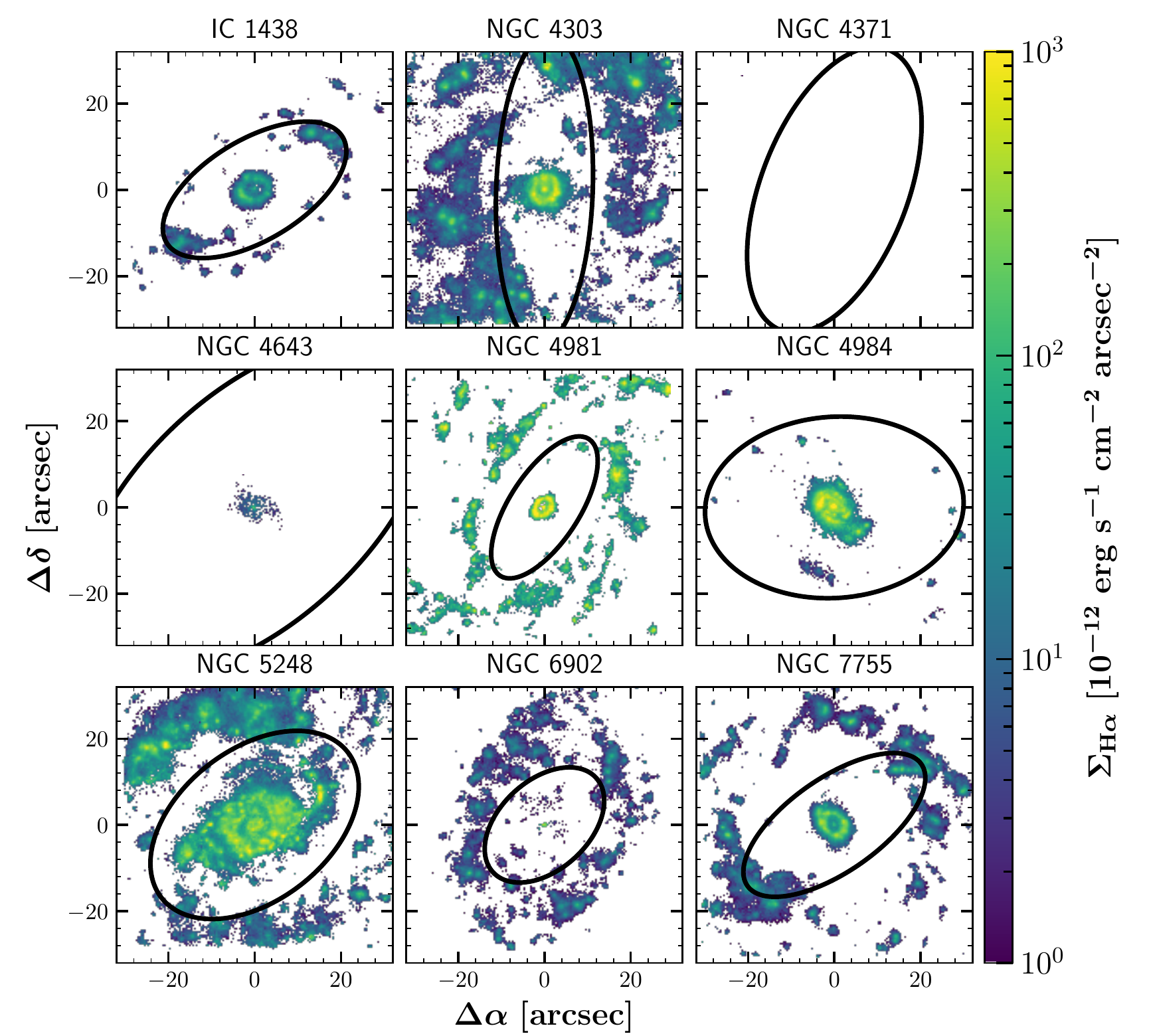}
	\caption{Dust-corrected H$\upalpha$ maps for the complete sample. Only fluxes with $\mathrm{S/N}_{\mathrm{H}\upalpha} > 5$ and $\mathrm{S/N}_{\mathrm{H}\upbeta} > 5$ are shown. Black ellipses outline the approximate extent of the galaxy bar. NGC 4371 does not have any H$\upalpha$ above the S/N cut.}
	\label{fig:sample_Ha}
\end{figure}

In Fig. \ref{fig:sample_Ha} we plot H$\upalpha$ maps for all galaxies in the sample. This figure shows that most galaxies have ongoing star formation either along spiral arms (NGC 4303, NGC 4981, NGC 5248), at the ends of the bar (IC 1438) or in a ring-like feature (NGC 6902, NGC 7755). Additionally, central regions often show nuclear structures (discs, rings or point sources) that are partially caused by star formation as well as by ionisation from the AGN (as revealed by the BPT diagrams).

That being mentioned, there is a clear lack of star formation between the centre and the ends of the bar for all bars. Some galaxies show star formation at the edges of the bar (NGC 4303 and a few blobs in IC 1438, NGC 4981, NGC 6902, NGC 7755) while others show none at all. This means that for most galaxies there is a supply of cold gas in the outer disc that either does not reach the bar region or the star formation is suppressed within the bar, for example, by means of strong velocity gradients. Interestingly, ionised gas is seen in the centre of all galaxies except NGC 4371, indicating that gas has been flown inwards. In fact, colour maps of the TIMER galaxies in figure 2 of \citetalias{Gadotti2019} show dust lanes in the bars in seven of our galaxies, which implies the presence of cold gas flows. Only NGC 4371 and NGC 4643 show no clear sign of gas in the bar. We will connect these results with the stellar population analysis in the next section.

The galaxy NGC 5248 is a peculiar case. Seen in H$\upalpha$, it seems to have a very large nuclear disc ($\sim \rm 1\,kpc$) with spiral-like features attached to it inside the bar region. It will show as an outlier in the subsequent plots in this paper.

\subsection{Stellar ages and metallicities}
\label{sect:age_z_maps}

\begin{figure}
	\centering
	\includegraphics[width=\columnwidth]{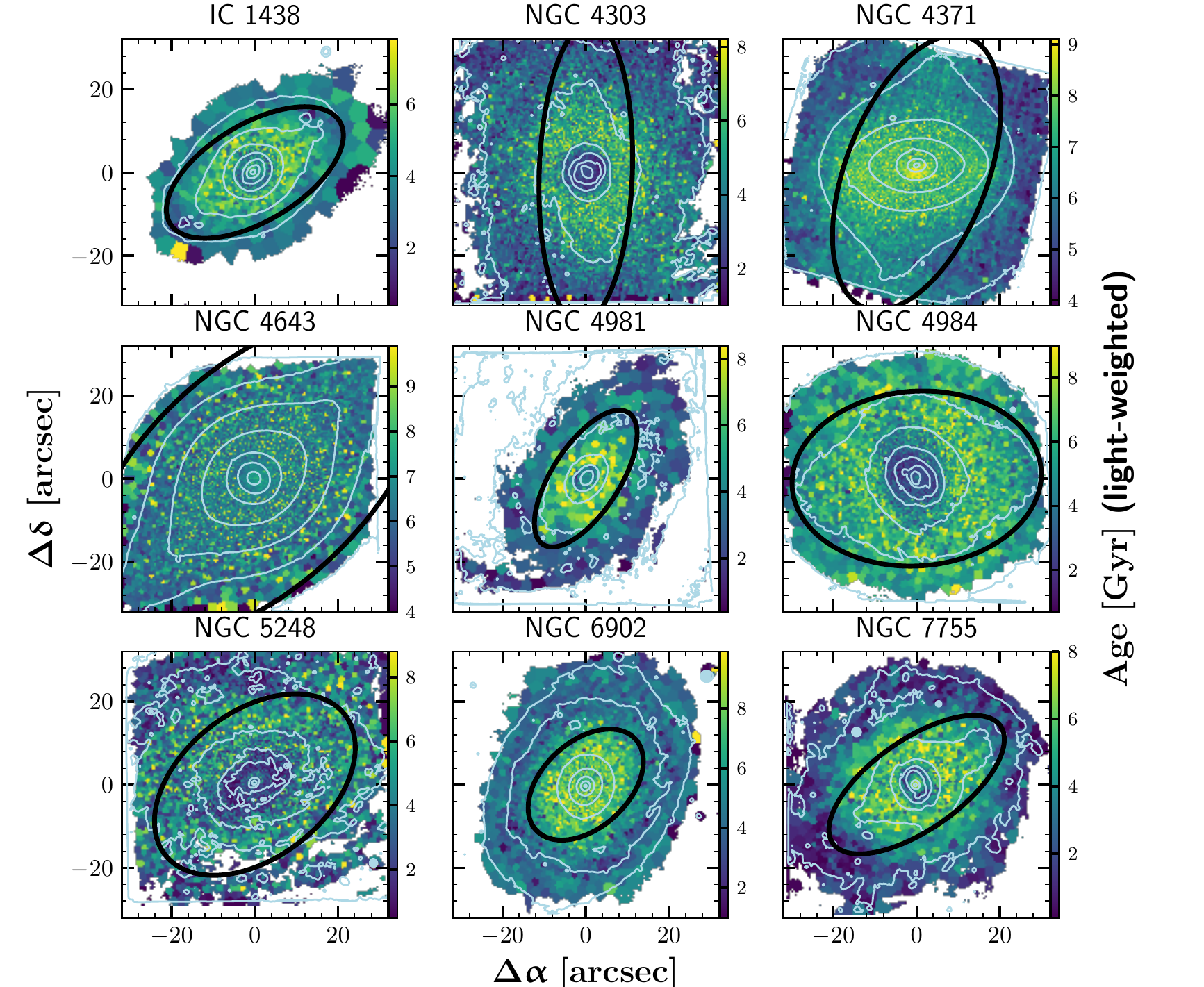}
	\caption{Light-weighted maps of mean stellar ages. Contours of the surface brightness distribution from the MUSE whitelight images are shown in white. Black ellipses outline the approximate extent of the bars.}
	\label{fig:age_lw}
\end{figure}

\begin{figure}
	\centering
	\includegraphics[width=\columnwidth]{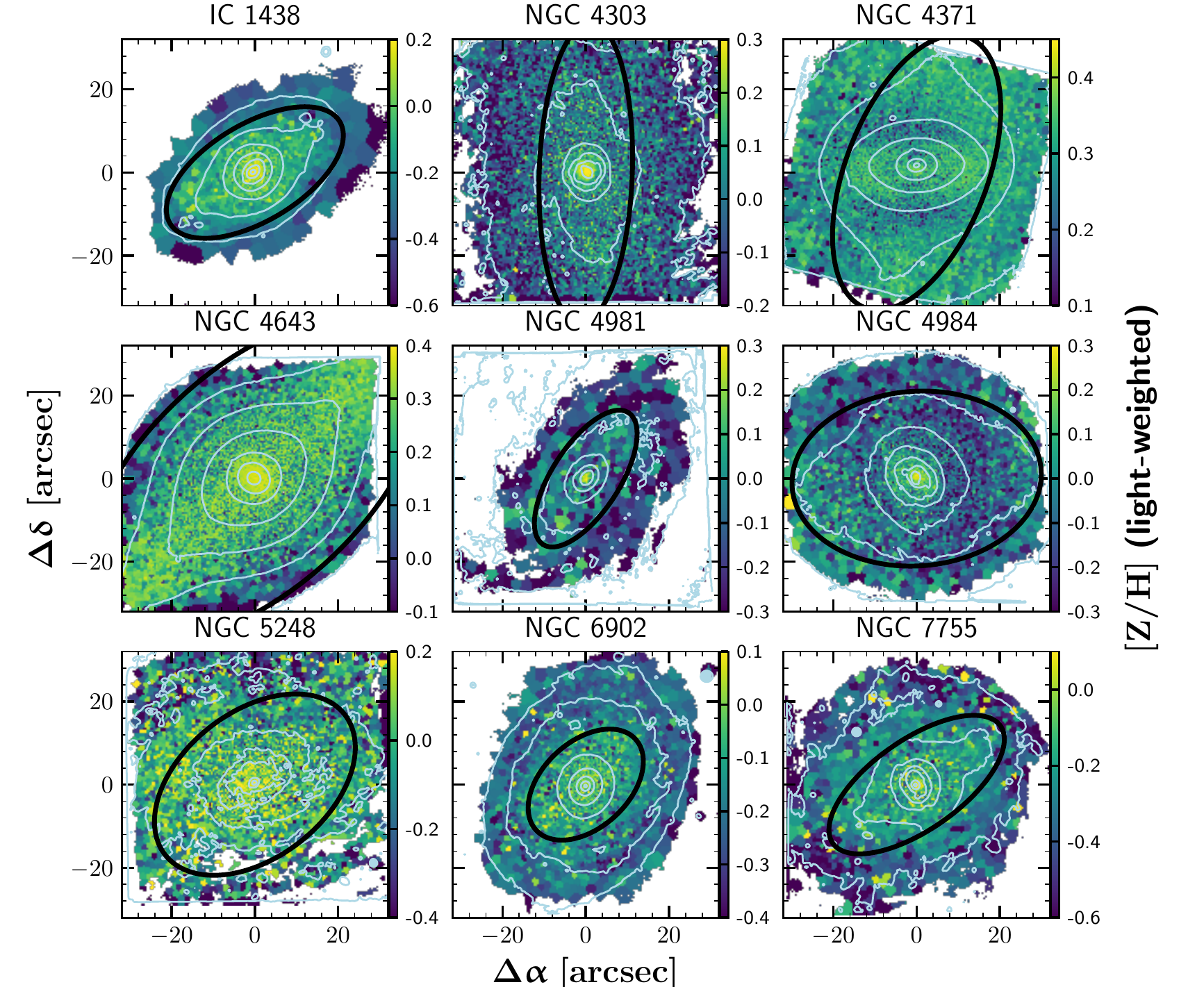}
	\caption{Light-weighted maps of mean stellar metallicities. Contours of the surface brightness distribution from the MUSE whitelight images are shown in white. Black ellipses outline the approximate extent of the bars.}
	\label{fig:z_lw}
\end{figure}

In Figs. \ref{fig:age_lw} and \ref{fig:z_lw}, we show spatially resolved maps of light-weighted mean stellar ages and metallicities, respectively. From a careful examination of the figures, we conclude that bars are typically older or as old as the part of the disc immediately surrounding the bar. This is in agreement with a suppression of star formation in the bars as seen in the previous section. Furthermore, we observe that bars are more metal-rich or as rich as the surroundings. Interestingly, for three galaxies (NGC 4371, NGC 4981, NGC 4984), we see low metallicity regions in the bar between the centre and the end of the bar. These regions of lower metallicities are seen more frequently in the mass-weighted maps (Fig. \ref{fig:age_z_mw}), where both bars and discs are mostly old. However, we caution that the conversion from light to mass is usually introducing additional uncertainties.

\subsection{[Mg/Fe] abundance ratios}
\label{sect:alphafe}

\begin{figure}
	\centering
	\includegraphics[width=\columnwidth]{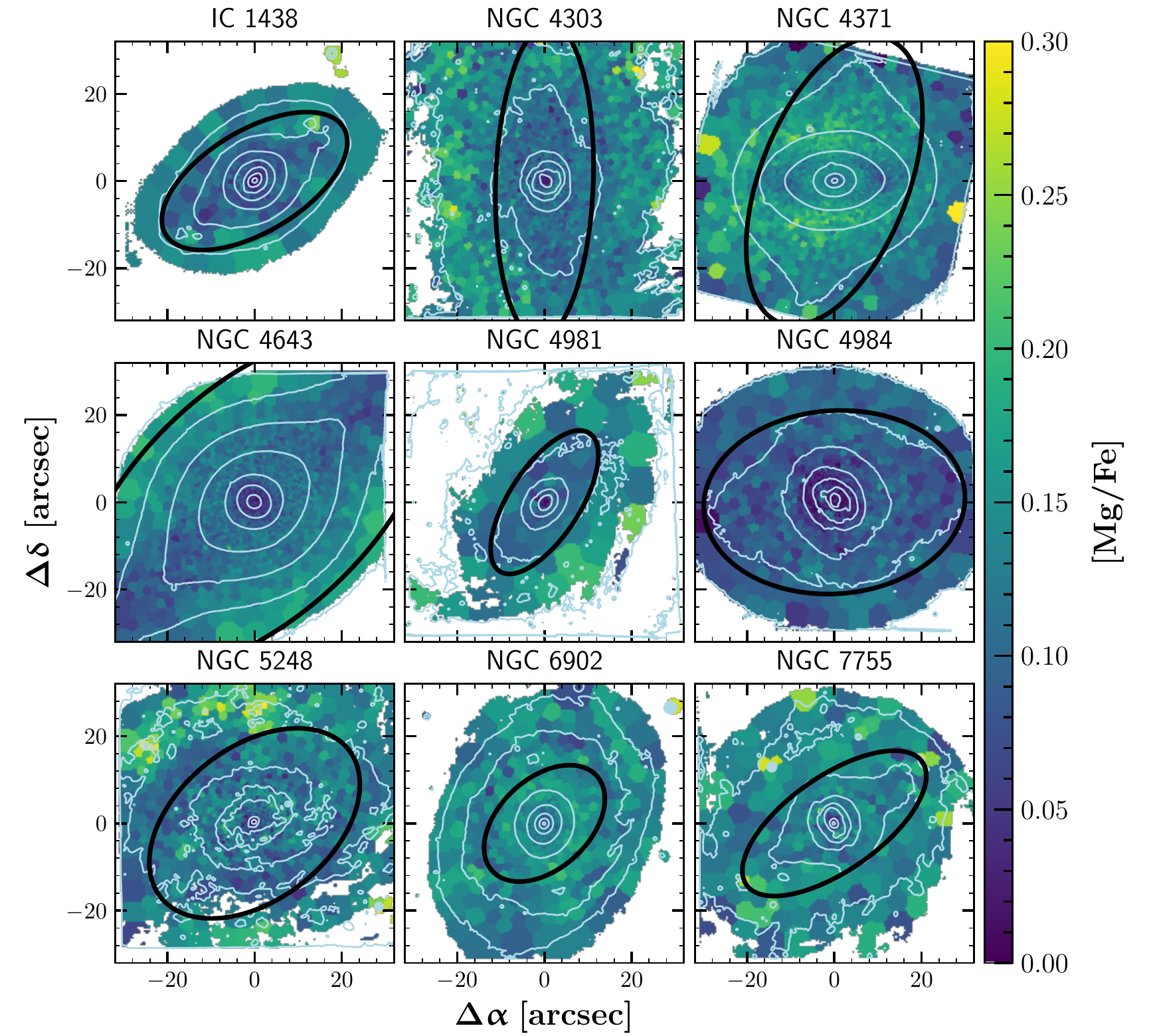}
	\caption{Spatially resolved Voronoi-binned maps of light-weighted [Mg/Fe]. Contours of the surface brightness distribution from the MUSE whitelight images are shown in white. The positions of the bars are shown in black, approximated by ellipses.}
	\label{fig:sample_alpha}
\end{figure}

The measurement of [Mg/Fe] can shed further light on the formation process of different components in a galaxy. This ratio has been traditionally used as a time-scale indicator of the SFH. One the one hand, Mg is almost exclusively produced by massive, exploding stars, and released to the interstellar medium in timescales of a few million years. On the other hand, the largest fraction of iron-peak elements are produced in type Ia supernovae (with a minor but important contribution from core-collapse supernovae, see e.g. \citealp{Maiolino2019} or \citealp{Bose2018}) which, after an episode of star formation, occur over an extended period of time following  a distribution of delay times \citep[e.g.][]{Matteucci1994, Greggio2008}. Quantifying the duration of the star formation using the [Mg/Fe] ratio is difficult, as the relation between these parameters can be modified by differences in the star formation rate, the initial mass function or the type Ia mechanisms. However, the comparison of [Mg/Fe] in different regions of the galaxies can give us a qualitative  idea about  the violence of the star formation processes and, therefore, about the physical mechanisms involved in their formation \citep[e.g.][]{Thomas1999}.

We present our measurements of [Mg/Fe] in Fig. \ref{fig:sample_alpha}. We notice that [Mg/Fe] in the bar is typically intermediate between the nuclear disc or nuclear ring component and the surrounding disc, where more elevated values of [Mg/Fe] are found. This is in accordance with the results found for inner bars as compared to the nuclear discs in the  double-barred galaxies NGC 1291 and NGC 5850 in the TIMER project \citep{Lorenzo-Caceres2019a} and supports the picture in which primary and inner bars are formed in similar ways.

Interestingly, by studying 16 barred galaxies from the Bars in Low Redshift Optical Galaxies (BaLROG) sample with IFU data from the SAURON instrument \citep{Bacon2001}, \citet{Seidel2016} found that the outer discs are less $\upalpha$-enhanced than the bars. However, the [$\upalpha$/Fe] of the discs in their sample is measured outside the bar radius. In contrast, the disc region that our measurements in TIMER cover is for most of the galaxies restricted to be within the radial range of the bar. This region, which encompasses the part of the disc that is within the bar radius but outside of the bar, is typically termed `star formation desert' \citep[SFD;][]{James2009,James2016}. In our sample, NGC 7755 in Fig. \ref{fig:sample_Ha} is a nice example of a SFD between the nuclear and the inner ring of H$\upalpha$. It seems that star formation is being suppressed by the bar in the SFD. In fact, a truncation of the SFH in SFDs has been found in observations \citep{James2016,James2018} and, as a more gradual decline, in cosmological zoom-in simulations \citep{Donohoe-Keyes2019}. In this work we, thus, find higher [Mg/Fe] abundances in the SFDs than in the bars. This result can be explained by a rapid suppression of star formation in the SFD after the formation of the bar and a more extended SFH in the bar. An even more extended period of star formation in the disc outside the radius of the bar, as reported by \citet{Seidel2016}, fits well within the same picture, in which many bars quench star formation within the bars themselves, while the outer discs are still forming stars \citep[e.g.][]{Neumann2019}.

\section{Stellar population gradients}
\label{sect:gradients}

\subsection{Profiles along the bar major and minor axis}
\label{sect:majorminor}

\begin{table*}
\caption{Gradients along the bar major and minor axis.}
\label{tbl:majorminor}
\centering

\resizebox{\textwidth}{!}{%
\begin{tabular}{l r r r r r r r r}
\hline\hline

\multicolumn{1}{c}{Galaxy} & $\Delta\,\mathrm{Age_{LW,MA}}$ & $\Delta\,\mathrm{Age_{LW,MI}}$ & $\Delta\,\mathrm{Age_{MW,MA}}$ & $\Delta\,\mathrm{Age_{MW,MI}}$ & $\Delta\,\mathrm{[Z/H]_{LW,MA}}$ & $\Delta\,\mathrm{[Z/H]_{LW,MI}}$ & $\Delta\,\mathrm{[Z/H]_{MW,MA}}$ & $\Delta\,\mathrm{[Z/H]_{MW,MI}}$\\
\multicolumn{1}{c}{(1)} & \multicolumn{1}{c}{(2)} & \multicolumn{1}{c}{(3)} & \multicolumn{1}{c}{(4)} & \multicolumn{1}{c}{(5)} & \multicolumn{1}{c}{(6)} & \multicolumn{1}{c}{(7)} & \multicolumn{1}{c}{(8)} & \multicolumn{1}{c}{(9)}\\

\hline
IC1438   &       $-1.44 \pm 0.17$        &       $-0.75 \pm 0.11$        &       $0.56 \pm 0.11$         &       $-1.10 \pm 0.59$        &       $-0.11 \pm 0.04$        &       $-0.18 \pm 0.02$        &       $0.01 \pm 0.02$         &       $-0.16 \pm 0.04$ \\
NGC4303  &       $-1.78 \pm 0.11$        &       $-1.53 \pm 0.15$        &       $-0.04 \pm 0.20$        &       $-0.47 \pm 0.36$        &       $-0.05 \pm 0.02$        &       $-0.07 \pm 0.02$        &       $0.00 \pm 0.04$         &       $0.04 \pm 0.02$ \\
NGC4371  &       $-0.32 \pm 0.06$        &       $-1.19 \pm 0.16$        &       $-0.21 \pm 0.03$        &       $-0.59 \pm 0.08$        &       $0.01 \pm 0.01$         &       $-0.02 \pm 0.01$        &       $0.01 \pm 0.01$         &       $-0.00 \pm 0.01$ \\
NGC4643  &       $-0.17 \pm 0.04$        &       $-0.02 \pm 0.06$        &       $-0.08 \pm 0.04$        &       $-0.06 \pm 0.14$        &       $0.01 \pm 0.01$         &       $-0.04 \pm 0.05$        &       $0.01 \pm 0.01$         &       $-0.04 \pm 0.05$ \\
NGC4981  &       $-2.65 \pm 0.31$        &       $-1.26 \pm 0.63$        &       $-0.54 \pm 0.73$        &       $-0.06 \pm 0.67$        &       $-0.02 \pm 0.05$        &       $-0.07 \pm 0.04$        &       $0.01 \pm 0.06$         &       $-0.03 \pm 0.08$ \\
NGC4984  &       $-0.56 \pm 0.12$        &       $0.21 \pm 0.29$         &       $0.21 \pm 0.10$         &       $1.00 \pm 0.40$         &       $0.05 \pm 0.02$         &       $0.04 \pm 0.04$         &       $0.08 \pm 0.02$         &       $0.03 \pm 0.04$ \\
$\rm NGC5248^\star$  &       $0.07 \pm 0.71$         &       $-0.40 \pm 0.22$        &       $-0.22 \pm 1.31$        &       $-0.70 \pm 0.40$        &       $-0.10 \pm 0.03$        &       $0.02 \pm 0.04$         &       $-0.09 \pm 0.07$        &       $-0.03 \pm 0.07$ \\
NGC6902  &       $-1.70 \pm 0.21$        &       $-1.42 \pm 0.24$        &       $0.10 \pm 0.18$         &       $0.25 \pm 0.64$         &       $-0.00 \pm 0.02$        &       $-0.06 \pm 0.01$        &       $0.06 \pm 0.02$         &       $0.03 \pm 0.04$ \\
NGC7755  &       $-1.12 \pm 0.13$        &       $-1.34 \pm 0.16$        &       $-0.38 \pm 0.26$        &       $-1.09 \pm 0.22$        &       $-0.05 \pm 0.03$        &       $-0.04 \pm 0.03$        &       $0.01 \pm 0.02$         &       $-0.01 \pm 0.03$ \\
\hline
Mean & $-1.22 \pm 0.79$ & $-0.91 \pm 0.62$ & $-0.05 \pm 0.32$ & $-0.27 \pm 0.66$ & $-0.02 \pm 0.05$ & $-0.05 \pm 0.06$ & $0.02 \pm 0.03$ & $-0.02 \pm 0.06$
\\
\hline
\end{tabular}}
\tablefoot{
Column (1) gives the name of the galaxy; columns (2)-(5) show the gradients of mean ages divided into light-weighted (LW) and mass-weighted (MW) values, as well as into the gradients along the bar major axis (MA) and the minor axis (MI); columns (6)-(10) show the gradients of mean metallicities in the same format as for the ages. The last row shows the mean values for all galaxies and the standard deviation between individual objects. The standard deviation of the mean is larger than the propagated error of the individual objects. $^\star$NGC 5248 is a clear outlier in this and following analyses and not included in the calculation of the means.}
\end{table*}

So far, most research on stellar populations in bars has focussed on profiles along the bar major axis and compared either to the disc \citep[e.g.][]{Sanchez-Blazquez2011,Fraser-McKelvie2019} or the bar minor axis \citep{Seidel2016}. Before we present our results of a different approach, we first show gradients of ages and metallicities along the bar major and minor axis for the sake of comparison with previous studies.

We extracted the major/minor axis profiles from pseudo slits of $2\arcsec$ width on top of the Voronoi-binned 2D mean age and metallicity maps for each galaxy. The observed distance along the axes ($R_\mathrm{obs}$) was deprojected to the plane of the galaxy ($R_\mathrm{gal}$) by applying the formula

\begin{equation}
R_\mathrm{gal} = R_\mathrm{obs} \sqrt{\frac{\sin^2(\Delta PA)}{\cos^2(i)} + \cos^2(\Delta PA)},
\end{equation}

where $i$ is the inclination of the disc and $\Delta PA = PA_\mathrm{disc}-PA_R$ is the difference between the position angles of the disc and the axis of $R_\mathrm{obs}$.

\begin{figure}
	\centering
	\resizebox{0.8\hsize}{!}{\includegraphics[trim=0 70 0 70,clip]{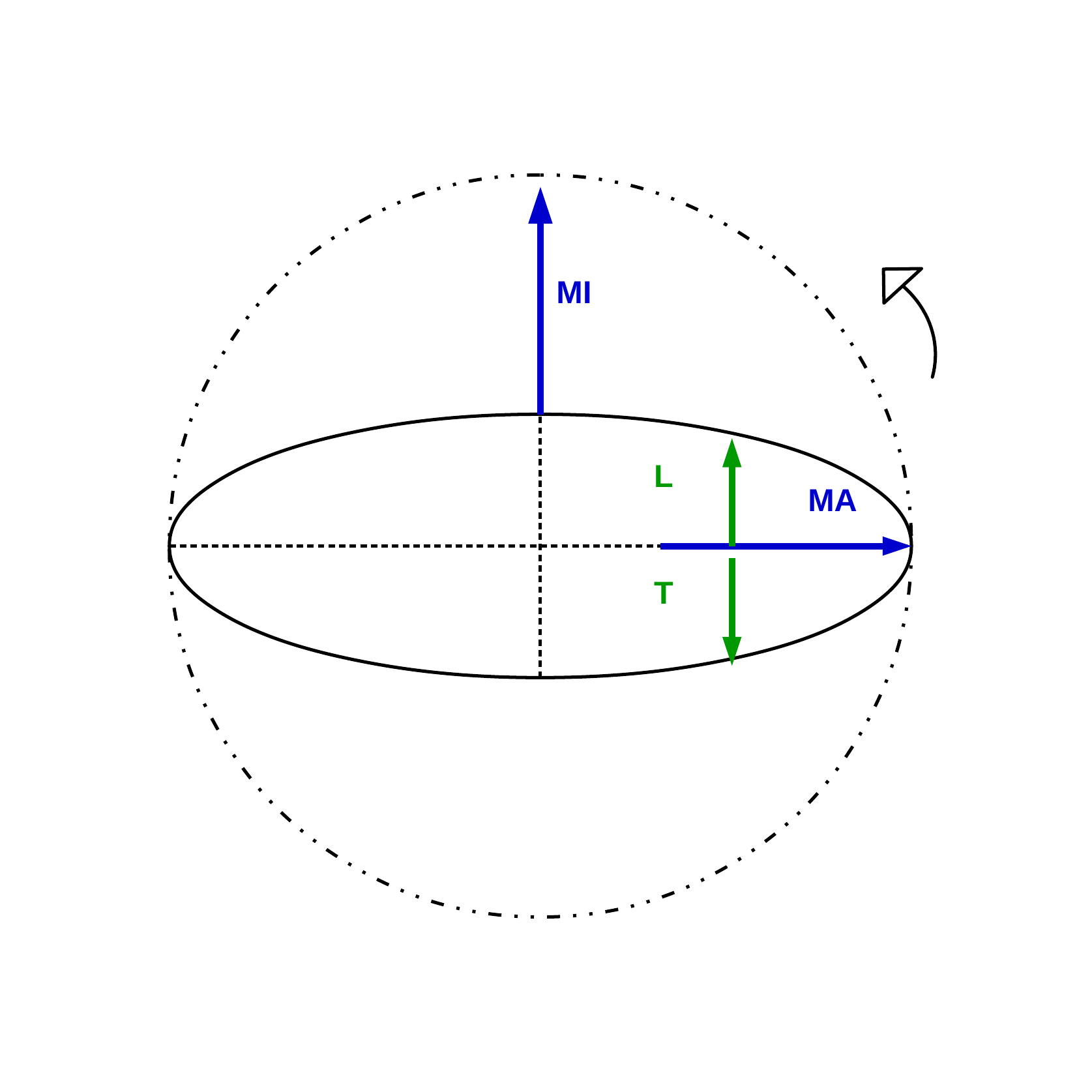}}
	\caption{Sketch of a rotating elongated bar in a face-on disc to illustrate the axes along which the stellar population gradients were measured. The solid-line elongated ellipse outlines the bar that rotates counter-clockwise as indicated by the arrow. The dashed lines mark the bar major and minor axes and the dot-dot-dashed circle gives the bar radius. The gradients discussed in Sect. \ref{sect:majorminor} are measured along the blue arrows and the gradients in Sect. \ref{sect:4cuts} along the green arrows. MA - along the major axis; MI - along the extension of the minor axis; L - from the major axis towards the leading edge of the bar; T - from the major axis towards the trailing edge of the bar.}
	\label{fig:sketch2}
\end{figure}

In Appendix \ref{apx:agemet_majorminor}, we explain the derivation of the gradients in more detail and show an example for the galaxy NGC 4303 in Figs. \ref{fig:majorminor_sketch} and \ref{fig:majorminor_profile}. Clear breaks in the major- and minor-axis profiles of age and metallicity are apparent in the inner regions of all galaxies, in agreement with \citet{Seidel2016}. These authors reported breaks commonly at $0.13 \pm 0.06$ bar length. We find two breaks which we visually identify: the first break is typically at or near the position of a nuclear structure such as a nuclear disc or nuclear ring; afterwards follows presumably a transition zone (between the regions where the nuclear structure/ the bar dominates the measurements) that ends at the second break. The second break in our sample is located on average at $r=0.29 \pm 0.09$ times the bar length. We chose the range between that break and the bar length to measure gradients using a linear regression along the bar major and minor axis. The implication is that what we call `gradient of the minor axis' is typically measured in the disc along the extension of the minor axis. This is illustrated schematically in Fig. \ref{fig:sketch2}. The gradients presented in the following are measured along the blue arrows annotated as \emph{MA} and \emph{MI}.

\begin{figure}
	\resizebox{\hsize}{!}{\includegraphics{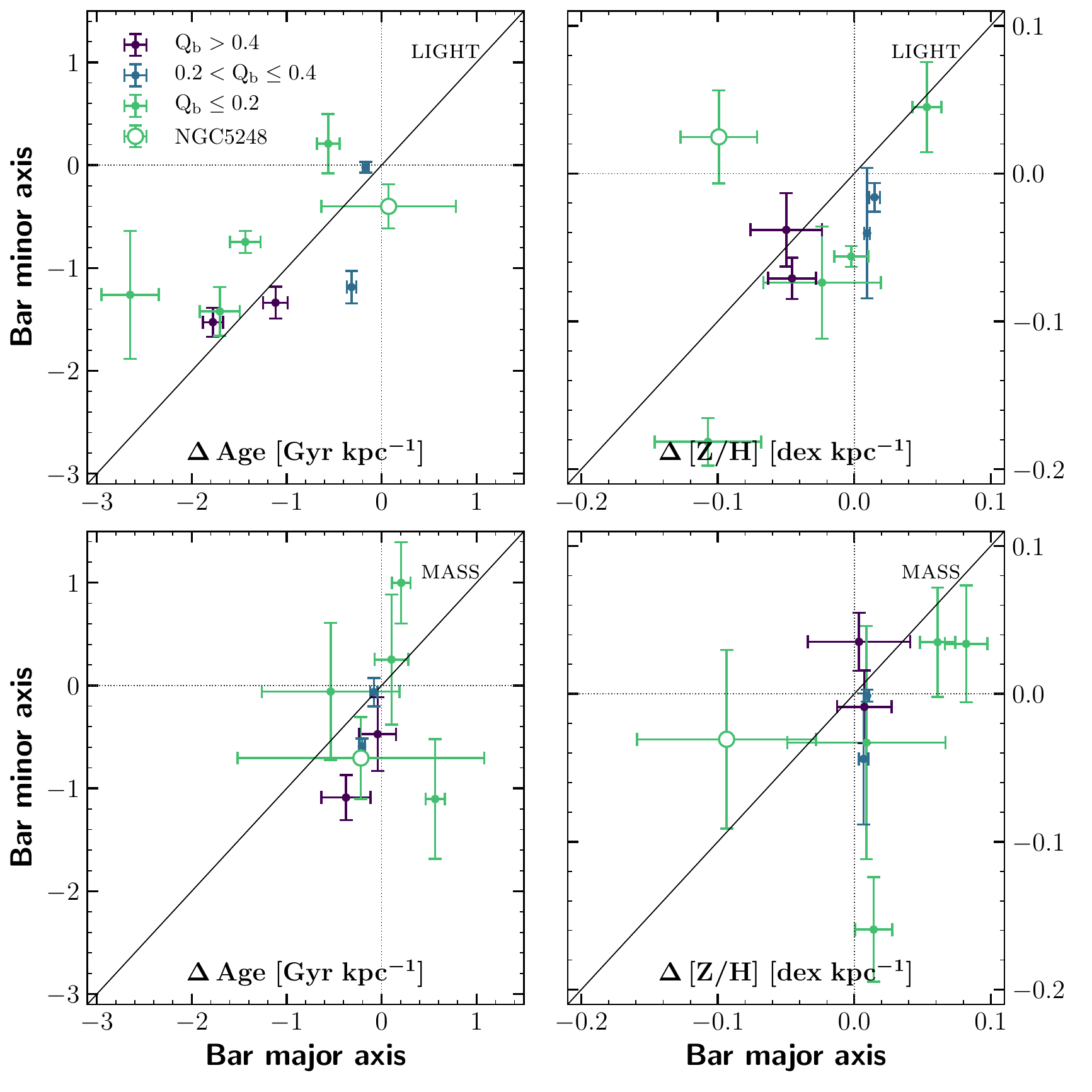}}
	\caption{Gradients of deprojected age and metallicity profiles for light- and mass-weighted mean values along the bar major and minor axes. An example of such profiles is shown in Fig. \ref{fig:majorminor_profile}. Gradients are measured between the second inner break and the bar length as seen in the aforementioned figure and discussed in the text. The minor axes are extended into the disc within the bar radius. Points are grouped into three bins of bar strength $Q_\mathrm{b}$. NGC 5248 is an outlier and shown separately. Error bars show standard errors from the linear regressions to the profiles.}
	\label{fig:majorminor_grad}
\end{figure}

In Fig. \ref{fig:majorminor_grad}, we present our results. The values are tabulated in Table \ref{tbl:majorminor}. Light-weighted age gradients are negative with no systematic difference between major and minor axes, yet on average the gradients are steeper on the major axis. Mass-weighted age gradients are flatter as compared to light-weighted ages with a larger scatter between individual objects on the minor axes than on the major axes. The mean mass-weighted values are $-0.05 \pm 0.32\ \mathrm{Gyr\ kpc^{-1}}$ for the major axes and $-0.27 \pm 0.66\ \mathrm{Gyr\ kpc^{-1}}$ for the minor axes (the errors are standard deviations from the mean). Hence, the age gradient along the major axis is on average flatter than on the minor axis.

Light-weighted gradients in [Z/H] are flatter on the major axes with respect to the minor axes and predominantly negative. They are slightly flatter for mass-weighted values with means of $+0.02 \pm 0.03\ \mathrm{dex\ kpc^{-1}}$ for the major axes and $-0.02 \pm 0.06\ \mathrm{dex\ kpc^{-1}}$ for the minor axes. Six out of nine objects have a positive and close-to-zero mass-weighted metallicity gradient along the major axis.

Additionally to comparing the gradients along the major and minor axes, we also separated them according to the bar strengths $Q_\mathrm{b}$ taken from \citet{Diaz-Garcia2016}. This parameter is a measure of the maximum gravitational torque in the bar region. It is commonly used as a criterion to classify bars into strong and weak. During the evolution of a bar, it usually grows longer and stronger \citep{Athanassoula2002,Elmegreen2007,Gadotti2011,Diaz-Garcia2016}. In Fig. \ref{fig:majorminor_grad} there is not much difference in the measured gradients between bars of different strengths. However, weaker bars seem to scatter more in the plot than stronger bars. Weak bars are usually smaller and less massive than strong bars. Consequently, the photometrical contrast with the underlying disc is smaller which could explain the larger scatter. A different explanation is that weaker bars could be younger and they may not have had enough time to `flatten' the gradients.

In general, our results that indicate a flattening along the bar major axis agree with recent results from the literature. In the BaLROG sample, \citet{Seidel2016} found that gradients of age, metallicity and [$\upalpha$/Fe] abundance along the bar major axis are flatter than the gradients along the minor axis, which are similar to those in discs of an unbarred control sample. In fact, they reported a metallicity gradient along the major axis of $0.03 \pm 0.07\ \mathrm{dex\ kpc^{-1}}$, very similar to our result. However, their mean gradient along the minor axis is $-0.20 \pm 0.04\ \mathrm{dex\ kpc^{-1}}$ and thus much steeper than the one we found in our sample. It is likely an effect of the different radial range over which the minor axis gradient was measured. Here, it is measured between the break at $\sim 0.29 \pm 0.09\ \times$ bar length and the full length of the bar. In contrast, they measured it mostly within the width of the bar. A flatter gradient along the bar was also confirmed in \citet{Fraser-McKelvie2019} studying 2D bar and disc regions of 128 strongly barred galaxies from the MaNGA survey \citep{Bundy2015}. Similarly, using long-slit observations, \citet{Sanchez-Blazquez2011} studied two of the bars in \citet{Perez2009} and found that they are flatter in age and metallicity as compared to the gradients along the major axis of the discs they are residing in.

In summary, we find close-to-zero mass-weighted age and metallicity gradients along the major axis of the bar that indicate the influence of the bar on the stellar populations. However, differences to the gradients along the minor axis are not very large and especially for weaker bars individual results produce significant scatter.

\subsection{Profiles across the width of the bar}
\label{sect:4cuts}

The stellar bar as we observe it in 2D projection is a superposition of stars that are trapped in mainly elongated orbits around the galaxy centre. Analyses of orbital structure in the gravitational potential of a barred disc galaxy reveal that bars are built from families of periodic and quasi-periodic orbits with different extents, elongations, and orientations \citep[e.g.][]{Contopoulos1980, Athanassoula1983, Pfenniger1984, Skokos2002, Skokos2002a}. One of these families is comprised of the $x_1$ orbits, which are elongated parallel to the bar major axis and build the backbone of the bar. Within the $x_1$ family, higher energy orbits are rounder and reach further into the disc and farther away from the bar major axis, whereas lower energy orbits are more elongated and closer to the bar major axis. Our aim is to investigate whether there are differences or trends in stellar populations across different orbits in the $x_1$ family that could help us to understand the formation and evolution of the bar.

\subsubsection{Selection of 1D pseudo-cuts}

\begin{figure}
	\resizebox{\hsize}{!}{\includegraphics{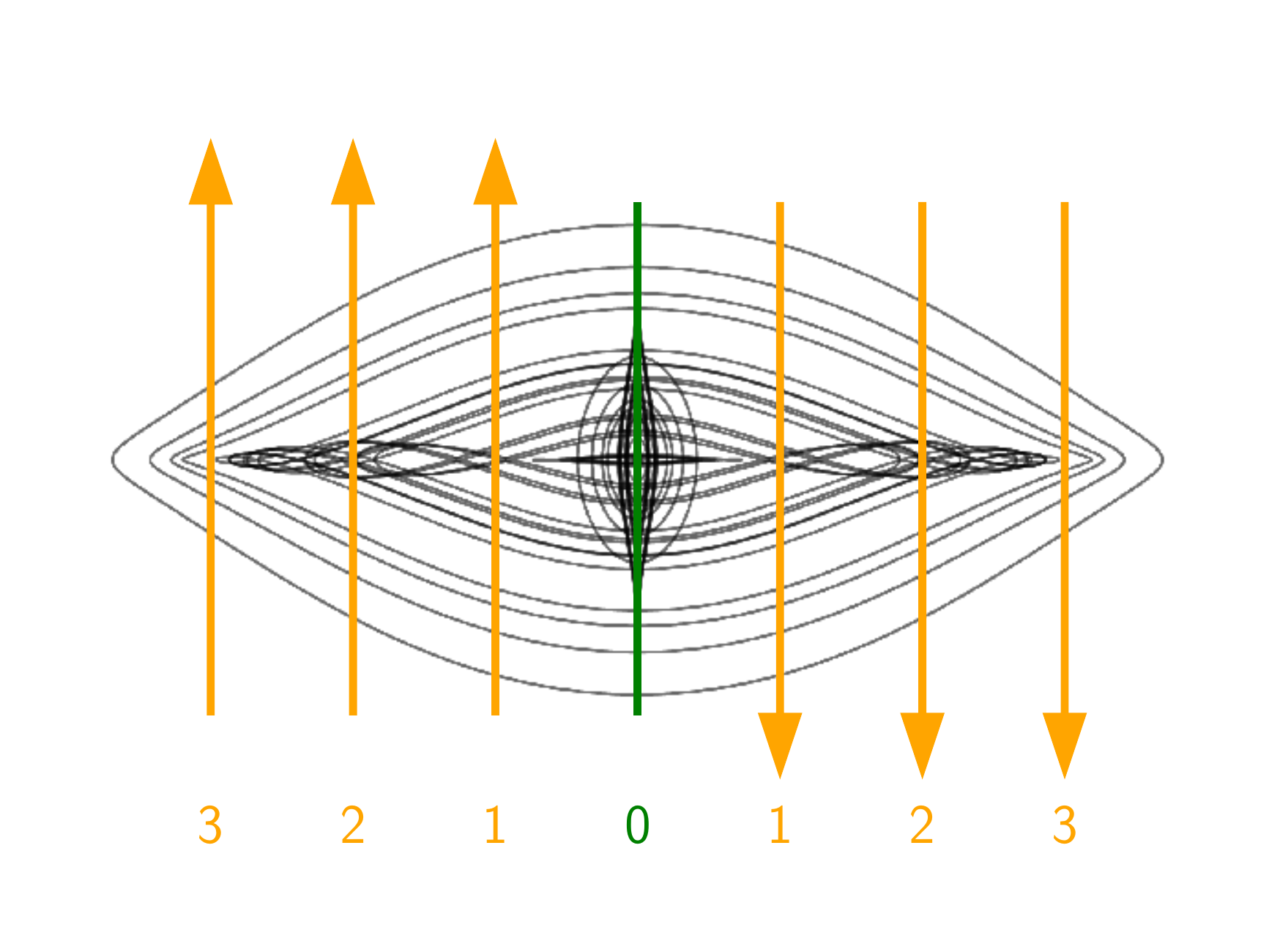}}
	\caption{Important closed periodic orbits in an analytic rotating bar potential. The bar major axis is horizontal. Orbits elongated parallel to the major axis are members of the $x_1$ family. Orbits elongated perpendicular to the major axis form the $x_2$ and $x_3$ families. Superimposed on the orbits, we show a sketch of the equidistant anti-parallel vectors along which the stellar populations are binned and averaged. The cuts are numbered from 0 to 3, where \#0 is crossing the centre of the galaxy and \#3 is at the end of the bar. These numbers will be used in the following figures and text.}
	\label{fig:sketch}
\end{figure}

We approached this problem by constructing a series of 1D cuts of $4\arcsec$ width perpendicular to the bar major axis\footnote{We note that the bar minor axis and the cuts are not perpendicular to the bar major axis in the projection on the sky, if the galaxy is not seen face-on. The angles are calculated such that they are orthogonal in the galaxy plane.}. We used four parallel cuts to both sides of the minor axis: a central cut on top of the minor axis, two cuts at the distances of one third and two thirds of the bar length, respectively, and one cut at the end of the bar. The cut at the end of the bar is not computed for the galaxies for which the complete length of the bar is not inside the FOV. Afterwards, every pair of equidistant cuts with respect to the minor axis was averaged in anti-parallel direction (with the exception of the central common cut). The procedure is illustrated in Fig. \ref{fig:sketch}. This approach ensures to average the leading edge of a rotating bar with the opposite leading edge, and the trailing edge with the opposite trailing edge.\footnote{The sense of rotation was determined assuming that spiral arms are trailing. For two galaxies, NGC 4371 and NGC 4643, we were not able to determine the sense of rotation due to the lack of spiral arm features.} The result is a set of four profiles going from the leading to the trailing edge cutting across the widths of the bar at different distances from the centre.

\begin{figure}
	\resizebox{\hsize}{!}{\includegraphics{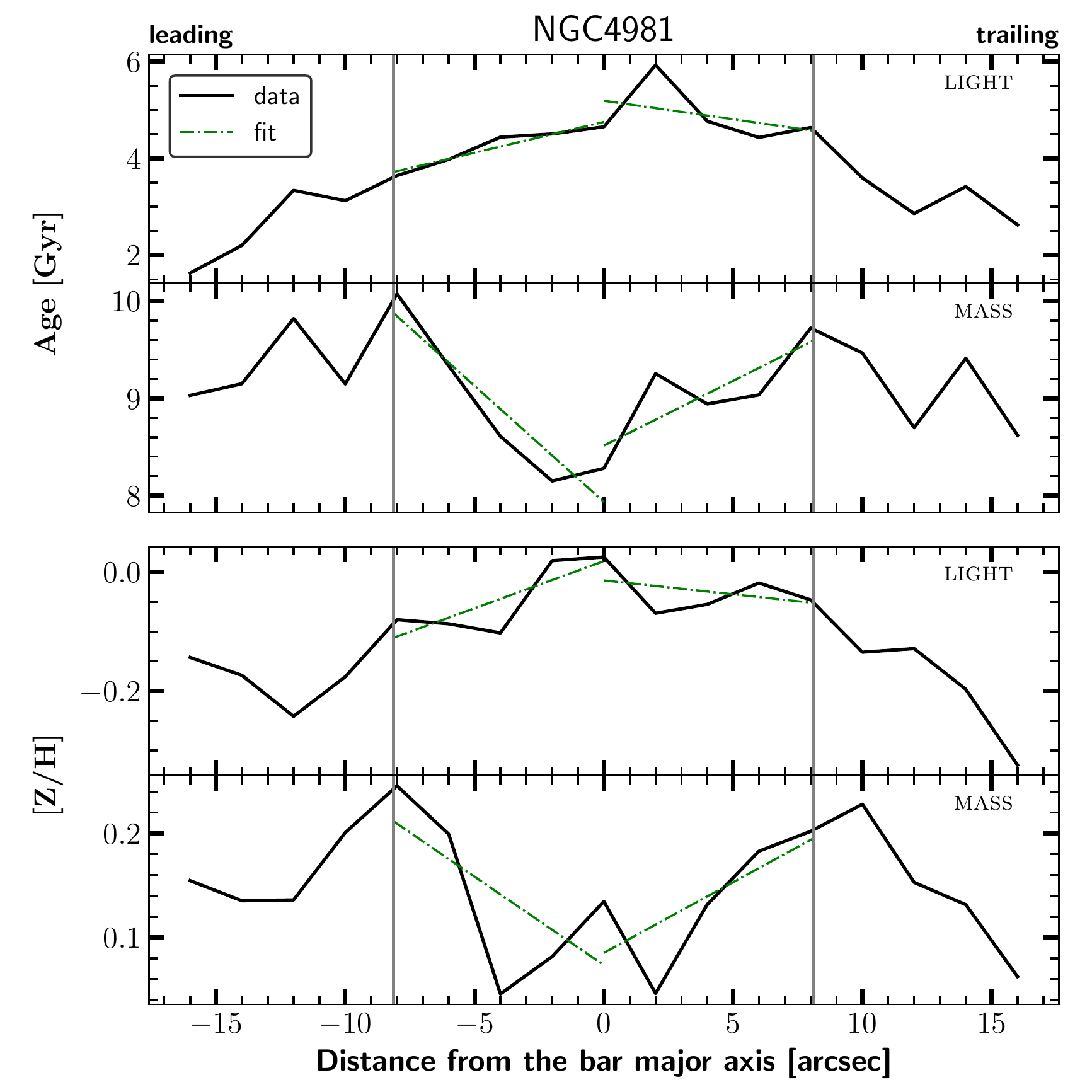}}
	\caption{Very simplified and shortened version of Figs. \ref{fig:AgeMet_first}-\ref{fig:AgeMet_last} as an example for NGC 4981. Shown are light- and mass-weighted age and metallicity profiles along cut \#2 (see Fig. \ref{fig:sketch}) perpendicular to the bar major axis. The width of the cuts is $4\arcsec$ and bins are equally spaced every $2\arcsec$ along the cuts. The two cuts \#2 are averaged in anti-parallel direction as shown in Fig. \ref{fig:sketch}. Vertical lines mark the edges of the bar. Linear regression fits from the major axis to the edges of the bar are shown in green. The slopes of these fits are tabulated in Table \ref{tbl:cuts} and plotted in Figs. \ref{fig:gradient_cuts_age} and \ref{fig:gradient_cuts_met}.}
	\label{fig:simplified_cut_profile}
\end{figure}

In Figs. \ref{fig:AgeMet_first}-\ref{fig:AgeMet_last} in Appendix \ref{apx:agemet_4cuts} we show the same 2D maps of light- and mass-weighted mean ages and metallicities for all galaxies but here together with the four aforementioned profiles. Additionally, along the cuts, we plot H$\alpha$ densities and total surface brightness. We also mark the position of dust lanes. A very simplified version of these figures is shown for the galaxy NGC 4981 in Fig. \ref{fig:simplified_cut_profile}. We will discuss our method and general results on the basis of this simplified example. For in-depth details of single objects, we refer the reader to Appendix \ref{apx:agemet_4cuts}.

The profiles in Fig. \ref{fig:simplified_cut_profile} show light- and mass-weighted mean age and metallicity along the averaged cut \#2 following the annotation in Fig. \ref{fig:sketch}, i.e. the cut at 2/3 of the bar length. Gradients were computed from linear regression fits to the profiles from the major axis towards both edges of the bar, which were determined from the lengths and ellipticites of the bar in Table \ref{tbl:sample} and \citet{Herrera-Endoqui2015}. For simplicity, in this method, the shape of the bar is assumed to be rectangular. The difference between this approach and the method described in Sect. \ref{sect:majorminor} can be seen in Fig. \ref{fig:sketch2}. The gradients described in this subsection are measured along the green arrows annotated as \emph{L} and \emph{T}. This procedure was done for all galaxies. We selected the cut at the distance of two thirds of the bar length to the centre because it is far enough away from the centre to be not contaminated by a central component and it is still well within the bar region.  

\subsubsection{Age and metallicity gradients}

\begin{figure}
	\resizebox{\hsize}{!}{\includegraphics{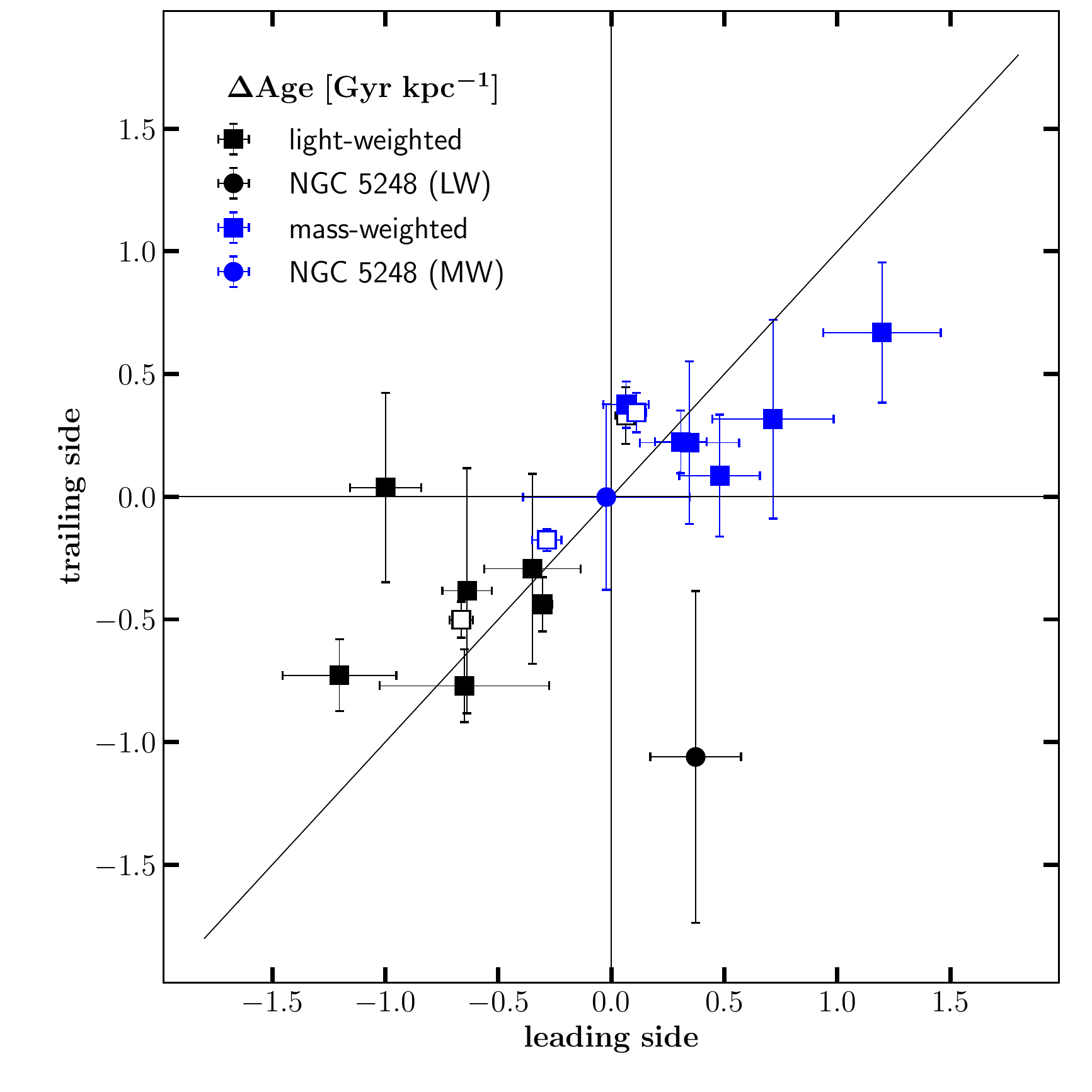}}
	\caption{Gradients of light- and mass-weighted mean stellar age profiles along the cut \#2 (see Fig. \ref{fig:sketch}) perpendicular to the bar major axes (at 2/3 of the bar length) as shown in Fig. \ref{fig:simplified_cut_profile}, as well as in Figs. \ref{fig:AgeMet_first}-\ref{fig:AgeMet_last} and described in the text. Shown are the gradients from the major axes towards the leading (x-axis) and trailing (y-axis) edges of the bars. The sense of rotation was determined assuming that spiral arms are trailing. Empty markers show galaxies for which the rotation is not clear. NGC 5248 is an outlier and shown separately. Error bars show standard errors from the linear regressions to the profiles.}
	\label{fig:gradient_cuts_age}
\end{figure}

\begin{figure}
	\resizebox{\hsize}{!}{\includegraphics{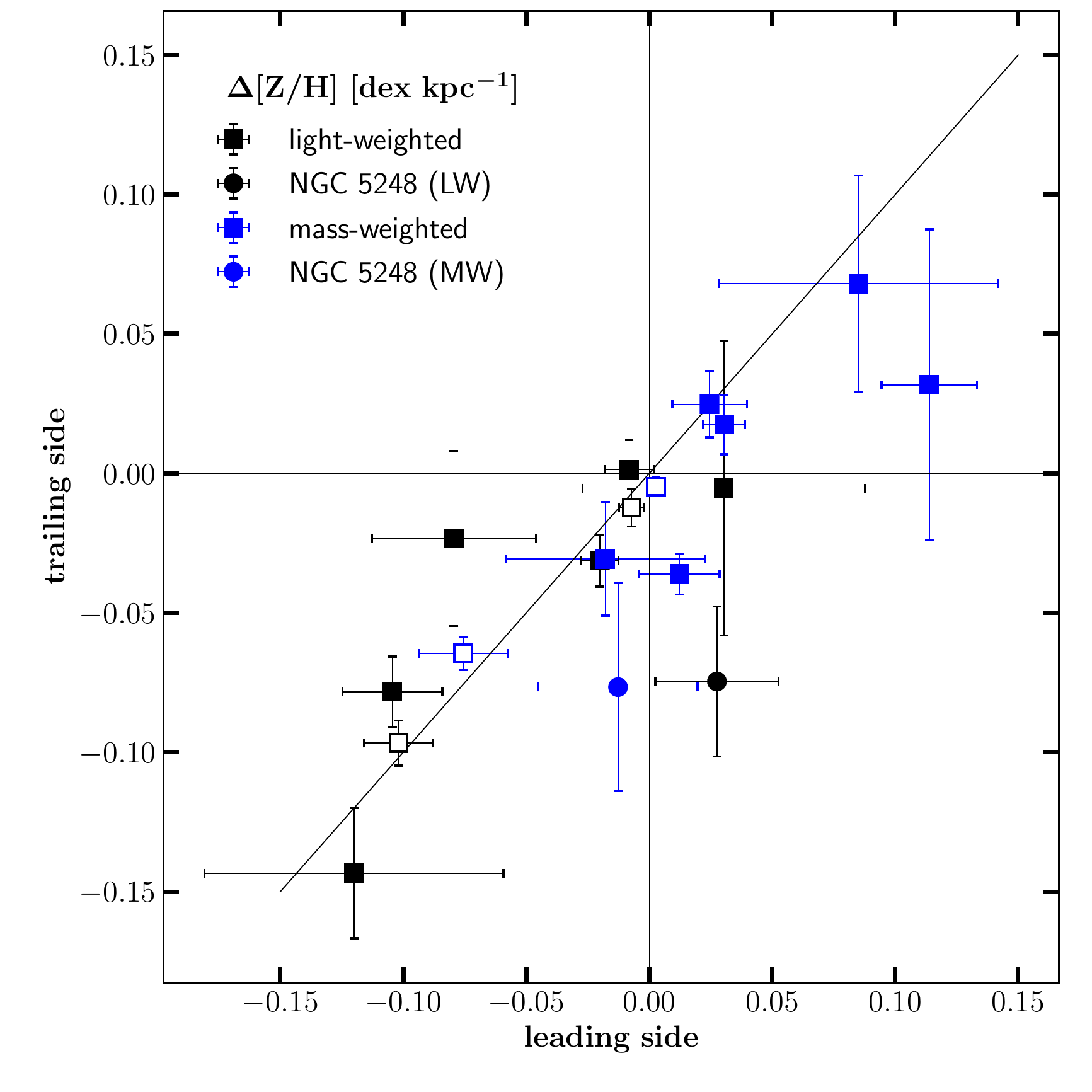}}
	\caption{Same as Fig. \ref{fig:gradient_cuts_age} but for gradients of mean stellar metallicities.}
	\label{fig:gradient_cuts_met}
\end{figure}

The results are presented in Fig. \ref{fig:gradient_cuts_age} for ages and Fig. \ref{fig:gradient_cuts_met} for metallicities. We compare the gradients from the leading to the trailing side as well as the light-weighted and mass-weighted quantities.

In the top panel of Fig. \ref{fig:simplified_cut_profile}, we observe that the light-weighted age profile is peaked close to the major axis and decreases towards both sides. Furthermore, we see that the gradient is steeper on the leading edge. Interestingly, in the mass-weighted profile, we see exactly the opposite trend. This behaviour is common to almost all galaxies in the sample. It is clearly apparent in Fig. \ref{fig:gradient_cuts_age} as almost all light-weighted gradients are negative and located in the bottom-left quadrant above the one-to-one line and mass-weighted gradient are located in the positive top-right quadrant. Differences between the centre and the edge of the bar can be small, especially in mass-weighted ages ($\sim 2\,$Gyr in Fig. \ref{fig:simplified_cut_profile}) but are trustable given the systematic trends across different bars as seen in Fig. \ref{fig:gradient_cuts_age}. The slightly steeper light-weighted gradients on the leading edges correlate with the appearance of H$\upalpha$ on the edges of those bars as can be seen in Figs. \ref{fig:AgeMet_first}-\ref{fig:AgeMet_last}. NGC 5248 is a clear outlier similarly to what has been seen in previous plots. 

The differences between light-weighted and mass-weighted mean stellar population parameters are simply explained by biases towards different underlying stellar populations. Light-weighted mean ages are biased towards the ages of the youngest and, therefore, more luminous stars. This effect is much reduced in mass-weighted quantities, which rather emphasise populations with intermediate to old ages. The negative light-weighted age gradients are clear indication for the presence of young stellar populations on the edges of the bar with a slight predominance on the leading side. This is in agreement with the general picture that, if there is star formation in a bar, it is preferentially happening on the leading side \citep[e.g.][]{Sheth2008,Neumann2019}, but this is the first time this is seen in the mean ages of stellar populations. At the same time, the positive mass-weighted gradients indicate that the stellar populations close to the bar major axis are younger than at the edges. This is an important result that will be strengthened by more details in the SFH and discussed in the following section.

Light-weighted metallicity gradients are on average shallow and negative with a mean of $\rm -0.05 \pm 0.05\ dex\ kpc^{-1}$ on both sides. Mass-weighted gradients are a bit flatter and mostly positive with a mean of $\rm 0.04 \pm 0.05\ dex\ kpc^{-1}$ on the leading side and $\rm 0.01 \pm 0.05\ dex\ kpc^{-1}$ on the trailing side. Thus, there are no significant differences between the leading and trailing edge.

\begin{table*}
\caption{Gradients along the cut \#2 (Fig. \ref{fig:sketch}) perpendicular to the major axis, i.e. at $\rm 2/3 \times\,bar length$ distance from the centre.}
\label{tbl:cuts}
\centering

\resizebox{\textwidth}{!}{%
\begin{tabular}{l r r r r r r r r}
\hline\hline

\multicolumn{1}{c}{Galaxy} & $\Delta\,\mathrm{Age_{LW,L}}$ & $\Delta\,\mathrm{Age_{LW,T}}$ & $\Delta\,\mathrm{Age_{MW,L}}$ & $\Delta\,\mathrm{Age_{MW,T}}$ & $\Delta\,\mathrm{[Z/H]_{LW,L}}$ & $\Delta\,\mathrm{[Z/H]_{LW,T}}$ & $\Delta\,\mathrm{[Z/H]_{MW,L}}$ & $\Delta\,\mathrm{[Z/H]_{MW,T}}$\\
\multicolumn{1}{c}{(1)} & \multicolumn{1}{c}{(2)} & \multicolumn{1}{c}{(3)} & \multicolumn{1}{c}{(4)} & \multicolumn{1}{c}{(5)} & \multicolumn{1}{c}{(6)} & \multicolumn{1}{c}{(7)} & \multicolumn{1}{c}{(8)} & \multicolumn{1}{c}{(9)}\\

\hline
IC1438	 &	 $-0.65 \pm 0.38$	 &	 $-0.77 \pm 0.15$	 &	 $0.48 \pm 0.18$	 &	 $0.09 \pm 0.25$	 &	 $-0.12 \pm 0.07$	 &	 $-0.14 \pm 0.03$	 &	 $0.01 \pm 0.02$	 &	 $-0.04 \pm 0.01$ \\
NGC4303	 &	 $-1.00 \pm 0.16$	 &	 $0.04 \pm 0.39$	 &	 $0.72 \pm 0.27$	 &	 $0.32 \pm 0.41$	 &	 $0.03 \pm 0.06$	 &	 $-0.01 \pm 0.06$	 &	 $0.11 \pm 0.02$	 &	 $0.03 \pm 0.06$ \\
$\rm NGC4371^\dagger$	 &	 $-0.66 \pm 0.06$	 &	 $-0.50 \pm 0.08$	 &	 $-0.29 \pm 0.07$	 &	 $-0.18 \pm 0.05$	 &	 $-0.01 \pm 0.01$	 &	 $-0.01 \pm 0.01$	 &	 $0.00 \pm 0.01$	 &	 $-0.00 \pm 0.01$ \\
$\rm NGC4643^\dagger$	 &       $0.06 \pm 0.05$         &       $0.33 \pm 0.12$         &       $0.11 \pm 0.05$         &       $0.34 \pm 0.09$         &       $-0.10 \pm 0.02$        &       $-0.10 \pm 0.01$        &       $-0.08 \pm 0.02$        &
       $-0.06 \pm 0.01$ \\
NGC4981	 &	 $-0.64 \pm 0.11$	 &	 $-0.38 \pm 0.50$	 &	 $1.20 \pm 0.26$	 &	 $0.67 \pm 0.29$	 &	 $-0.08 \pm 0.04$	 &	 $-0.02 \pm 0.04$	 &	 $0.09 \pm 0.06$	 &	 $0.07 \pm 0.04$ \\
NGC4984	 &	 $-0.30 \pm 0.05$	 &	 $-0.44 \pm 0.12$	 &	 $0.31 \pm 0.12$	 &	 $0.22 \pm 0.13$	 &	 $-0.01 \pm 0.02$	 &	 $0.00 \pm 0.02$	 &	 $0.02 \pm 0.02$	 &	 $0.02 \pm 0.02$ \\
$\rm NGC5248^{\star}$	 &	 $0.37 \pm 0.21$	 &	 $-1.06 \pm 0.68$	 &	 $-0.02 \pm 0.37$	 &	 $-0.00 \pm 0.38$	 &	 $0.03 \pm 0.03$	 &	 $-0.07 \pm 0.03$	 &	 $-0.01 \pm 0.04$	 &	 $-0.08 \pm 0.04$ \\
NGC6902	 &	 $-1.20 \pm 0.26$	 &	 $-0.73 \pm 0.15$	 &	 $0.07 \pm 0.10$	 &	 $0.38 \pm 0.10$	 &	 $-0.02 \pm 0.01$	 &	 $-0.03 \pm 0.01$	 &	 $0.03 \pm 0.01$	 &	 $0.02 \pm 0.02$ \\
NGC7755	 &	 $-0.35 \pm 0.22$	 &	 $-0.29 \pm 0.39$	 &	 $0.35 \pm 0.22$	 &	 $0.22 \pm 0.34$	 &	 $-0.10 \pm 0.03$	 &	 $-0.08 \pm 0.02$	 &	 $-0.02 \pm 0.05$	 &	 $-0.03 \pm 0.03$ \\

\hline
Mean & $-0.69 \pm 0.32$ & $-0.43 \pm 0.27$ & $0.52 \pm 0.36$ & $0.32 \pm 0.18$ & $-0.05 \pm 0.05$ & $-0.05 \pm 0.05$ & $0.04 \pm 0.04$ & $0.01 \pm 0.04$ 
\\
\hline
\end{tabular}}
\tablefoot{
Column (1) gives the name of the galaxy; columns (2)-(5) show the gradients of mean ages divided into light-weighted (LW) and mass-weighted (MW) values, as well as into the gradients along the leading edge of the bar (L) and the trailing edge (T); columns (6)-(10) show the gradients of mean metallicities in the same format as for the ages. The last row shows the mean values from all galaxies except NGC 4371, NGC 4643 and NGC 5248, and the standard deviation between individual objects. The standard deviation of the mean is larger than the propagated error of the individual objects. $^\dagger$For NGC 4371 and NGC 4643, we were unable to define the sense of rotation. For these galaxies, the sides (L/T) are arbitrarily chosen. $^{\star}$NGC 5248 is a clear outlier and not included in the calculation of the means.}
\end{table*}

\subsection{An alternative visualisation}
\label{sect:alternative_vis}

\begin{figure}
	\centering
	\includegraphics[width=\columnwidth]{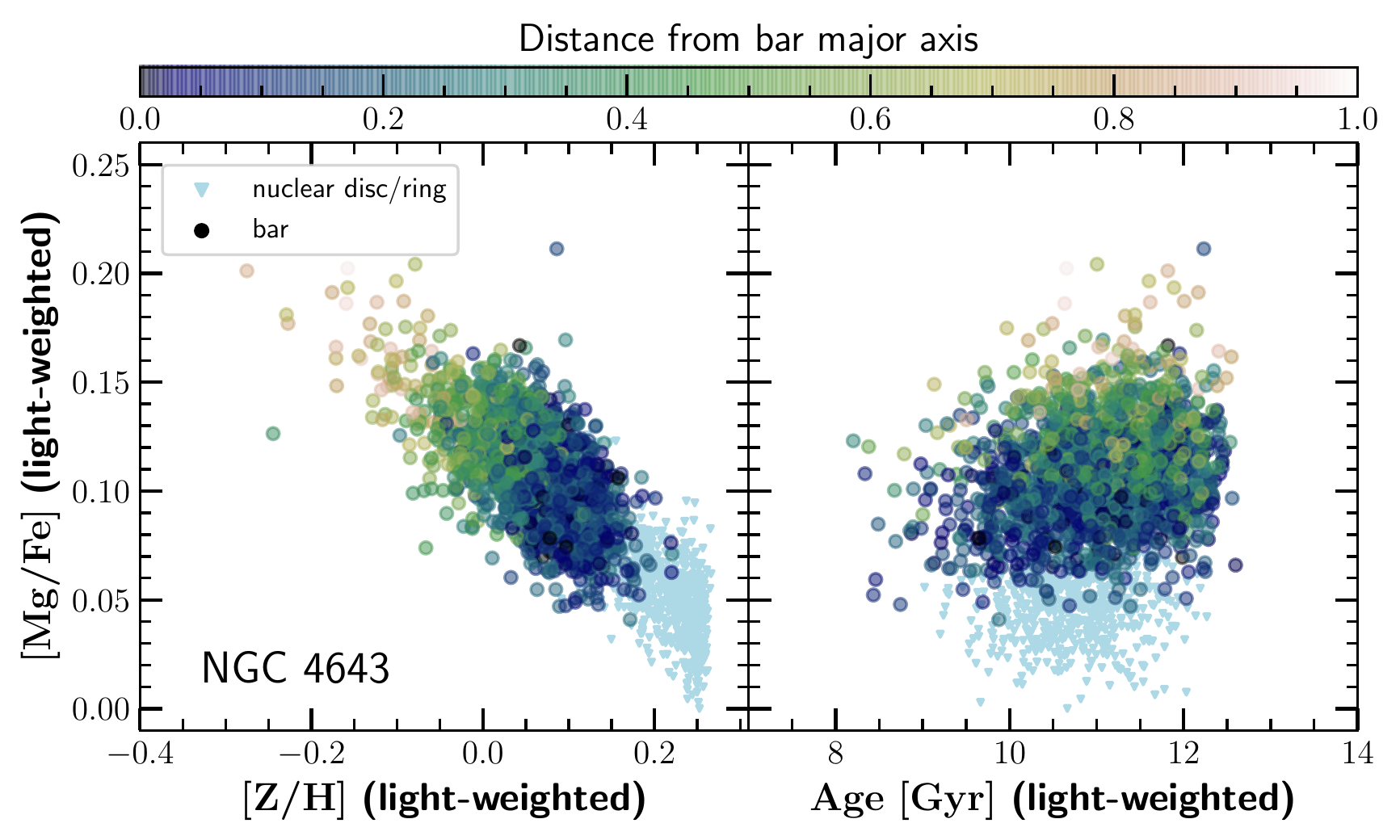}
	\caption{Comparison between light-weighted [Mg/Fe], metallicity and age of the stellar populations in the bar of NGC 4643. Each point corresponds to one Voronoi bin as shown in Fig. \ref{fig:sample_alpha}. Bins are coloured by their distance to the bar major axis and only bins within the bar are shown. Bins within the nuclear structure are marked with different symbols and in light blue. These plots are shown for all galaxies in Figs. \ref{fig:ref_age_z_lw} and \ref{fig:ref_age_z_mw}.}
	\label{fig:ref_NGC4643}
\end{figure}

An alternative way of visualising the data and results that we discussed in Sects. \ref{sect:maps} and \ref{sect:4cuts} is shown as an example for NGC 4643 in Fig. \ref{fig:ref_NGC4643} light-weighted and for the whole sample in Figs. \ref{fig:ref_age_z_lw} and \ref{fig:ref_age_z_mw} light-weighted and mass-weighted, respectively. In these plots we directly compare [Mg/Fe] with metallicity and age for each bin within the bar. The points are colour-coded by their shortest distance to the bar major axis normalised by the width of the bar. Bins that are within the nuclear structure close to the centre of the galaxy are shown as different symbols and in light blue. The radius of the nuclear structure is taken from Gadotti et al. (in prep.). The data presented in these figures is based on the analysis with \textls{\sc pPXF}, since \textls{\sc steckmap} does not provide [Mg/Fe] measurements. Notwithstanding, we remark again that \textls{\sc pPXF} and \textls{\sc steckmap} return analogous results for these galaxies (Bittner et al., in prep.).

The left panel in Fig. \ref{fig:ref_NGC4643} shows some clear trends for NGC 4643 that agree with our previous analysis. As we move away from the bar major axis the stars become more [Mg/Fe]-enhanced and more metal-poor. In the right panel we see that the stars in this particular bar are predominantly old, even in light-weighted mean ages that are biased towards younger populations. There is a small trend to older ages as the distance to the major axis increases, but the scatter is large due to the difficulty to separate old populations in the fit. These results agree with the positive age gradient and the negative metallicity gradient perpendicular to the bar major axis tabulated in Table \ref{tbl:cuts}.

These trends are observed for the majority of bars as it can be seen in Figs. \ref{fig:ref_age_z_lw} and \ref{fig:ref_age_z_mw}. However, it has to be noted that these figures mix all stars in the bar (except the nuclear structure) and they only separate stars perpendicular but not along the bar major axis. Thus, some of the trends that we observed along the clear cuts in Sect. \ref{sect:4cuts} might be washed out in particular cases in the figures presented here.

\section{Star formation histories}
\label{sect:SFH}

One single observation with the MUSE instrument of a galaxy provides 90\,000 spectra each of which contains information that makes it possible to disentangle, inter alia, the composite of young and old stellar populations, as well as metal-poor and metal-rich. The presentation of the full wealth of information from the spatially resolved SFH of a galaxy is a multi-dimensional problem and it is a challenge to best illustrate important aspects. Two-dimensional maps of mean ages and metallicities, as shown in Sect. \ref{sect:maps}, are projections that keep the spatial information but average the parameters along the axis of time. In this subsection, we present how stars of different ages shape the stellar bars that we observe. In the figures of SFH, we will use the same spatial binning scheme along the cuts perpendicular to the bar major axis as presented previously.

\begin{figure}
	\resizebox{\hsize}{!}{\includegraphics{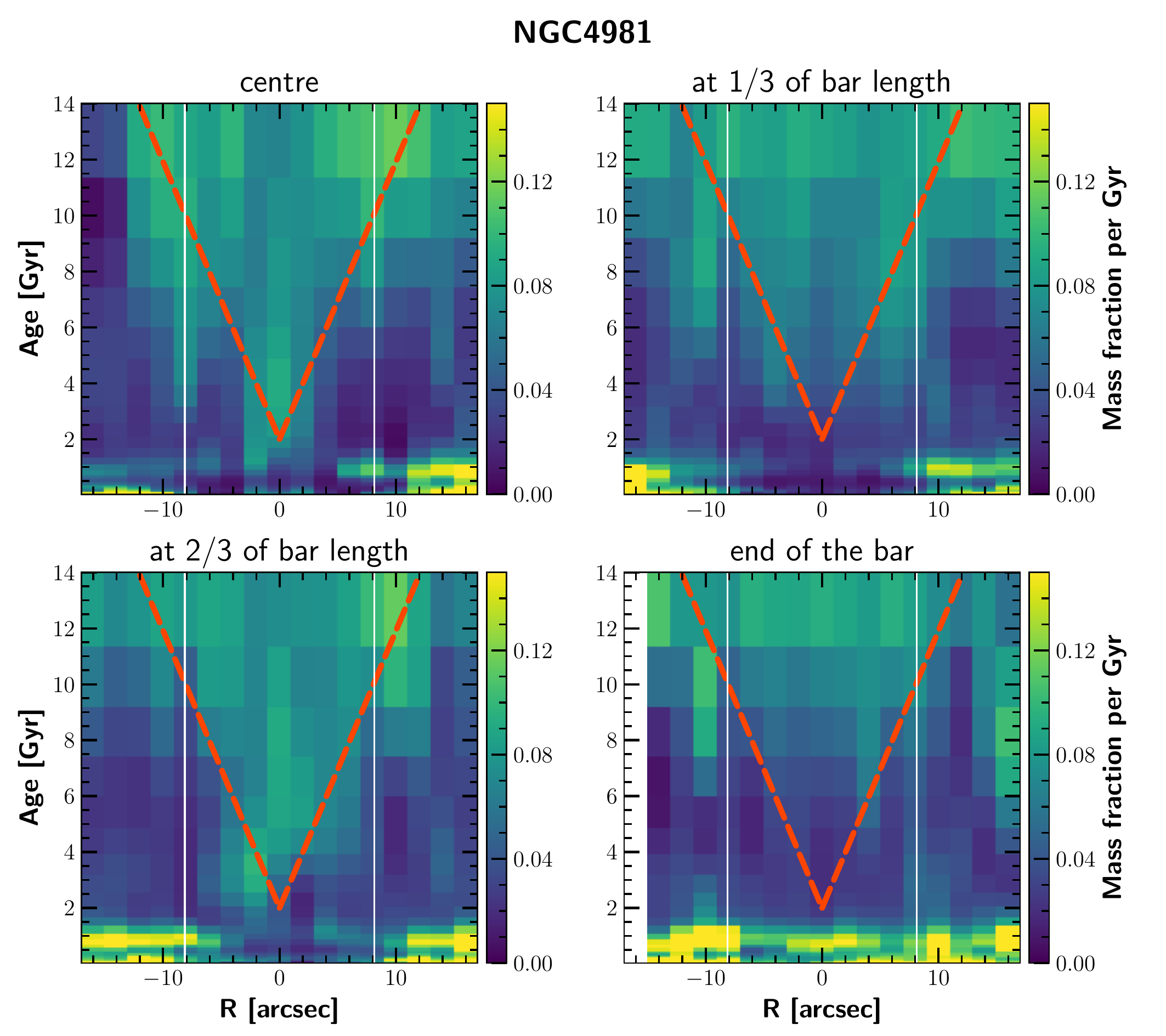}}
	\caption{SFHs of NGC 4981 along the cuts illustrated in the sketch in Fig. \ref{fig:sketch}. and in the Figs. \ref{fig:AgeMet_first}-\ref{fig:AgeMet_last}. Each panel shows one of the four cuts. The x-axis shows the distance to the bar major axis along the cut and the y-axis the age of the population. Colour-coded is the fraction of stellar mass that is at a certain position and was formed at a certain time. The mass is normalised within each spatial bin (vertically in this diagram). The mass fractions are further divided by the size of the bin on the y-axis to correct for non-equally spaced bins on the linear age axis. White vertical lines mark the edges of the bar. Red lines are plotted on top to indicate the `V-shape' discussed in the text.}
	\label{fig:SFH_NGC4981_mass}
\end{figure}

The SFHs are shown as an example for NGC 4981 along the four cuts in four different panels in Fig. \ref{fig:SFH_NGC4981_mass}. The last panel shows the profile at the end of the bar. In this panel, we see a very young and an old population with not much variation across the cut, which highlights that there is not much difference between the ends of the bar and the disc. In the H$\upalpha$ maps in Fig. \ref{fig:sample_Ha}, in fact, we see ongoing star formation at the end of the bar in NGC 4981, in agreement with the very young populations seen in the SFH. We now address the second and third panel, both of which contain information with less contamination from the nuclear structure (first panel) and the outer disc (last panel). The plots present clear evidence of a very young stellar population in the main disc, here seen as bright features of less than 1$\,$Gyr left and right to the edges of the bar. Additionally, we recognise a `V-shape' in the ages above $2\,$Gyr, where stars at intermediate ages between 2-8$\,$Gyr are more concentrated close to the major axis, while the oldest population ($> 8\,$Gyr) is spread across the whole spatial range. This feature is not exclusive for this galaxy but can be seen in at least 5 out of 9 galaxies in our sample (IC 1438, NGC 4643, NGC 4981, NGC 6902, NGC 7755). Plots of SFHs for the complete set of galaxies can be found in Fig. \ref{fig:SFH_all} in Appendix \ref{apx:SFH}. A consequence of this `V-shape' structure in the SFH is a positive age gradient from the major axis towards the edges of the bar that we observed indeed for all but one galaxy in the mass-weighted mean ages in Fig. \ref{fig:gradient_cuts_age}.

These results indicate that intermediate age stellar populations are concentrated on more elongated orbits closer to the bar major axis than older stellar populations. They are consistent with the findings from idealised thin (kinematically cold/young) plus thick (kinematically hot/old) disc $N$-body galaxy simulations in \citet{Fragkoudi2017}. In their figure 2, they show that the colder component forms a strong and thin bar, while the hotter component forms a weaker and rounder bar \citep[see also][]{Wozniak2007,Athanassoula2017,Debattista2017,Fragkoudi2018}. We explore the parallels between our results from observations with simulations further in Sect. \ref{sect:auriga}.

\section{Discussion}
\label{sect:discussion}

\subsection{The origin of the V-shaped age distribution: input from cosmological simulations}
\label{sect:auriga}

\begin{figure}
	\resizebox{\hsize}{!}{\includegraphics{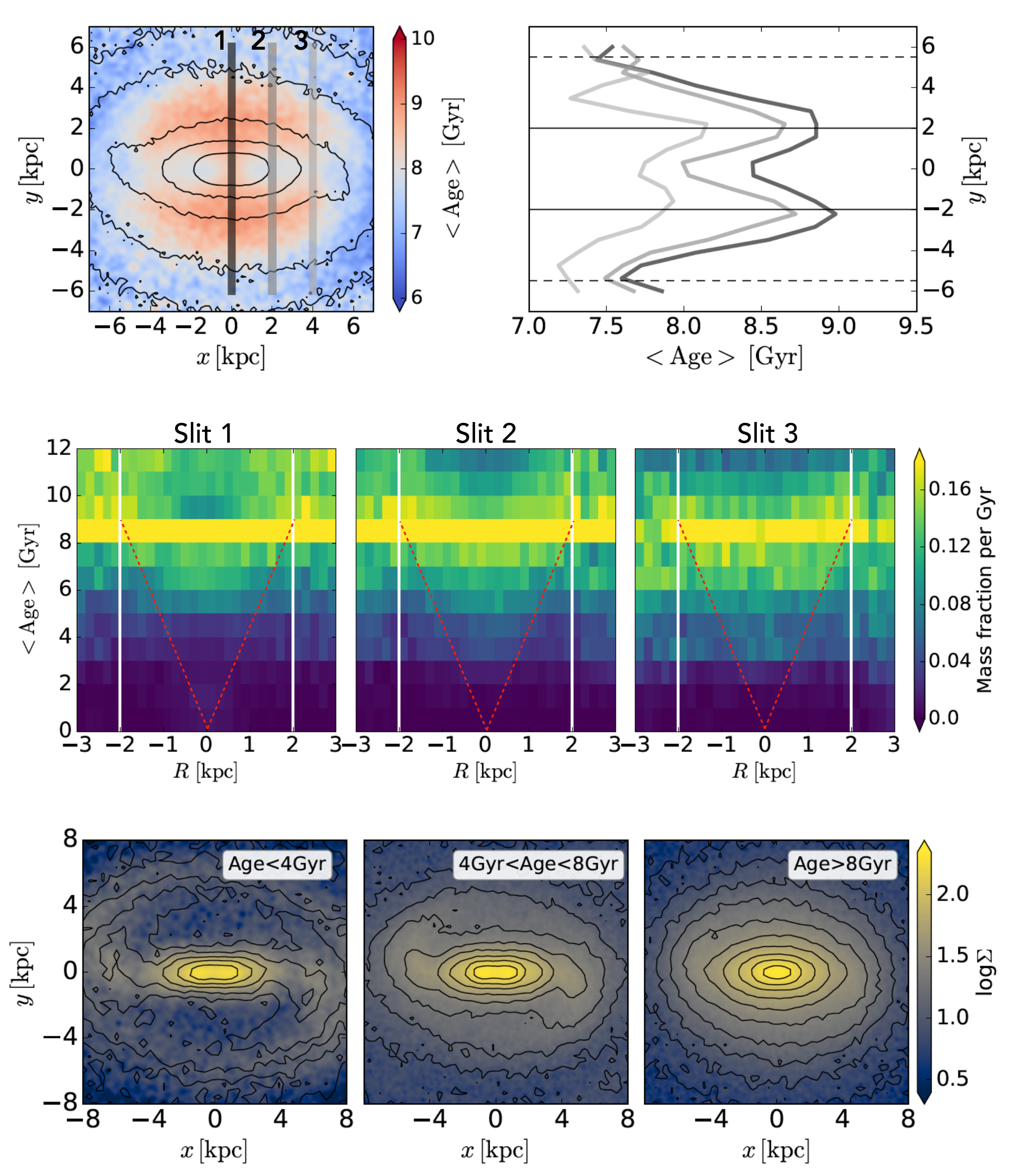}}
	\caption{Results from Auriga simulations. The top-left panel shows a face-on mass-weighted age map of halo 18 from the Auriga sample, over-plotted with three vertical lines that indicate the cuts to extract the data shown in the other panels. The bar major axis is horizontal. The top-right panel shows the mean stellar age profiles along the cuts. The middle row presents SFHs diagrams for each of the cuts (the left panel corresponds to the dark slit and the rightmost panel corresponds to the light grey slit). The axis and the colour-coding are the same as in our observations shown in Fig. \ref{fig:SFH_NGC4981_mass}. The white solid lines mark the edges of the bar and the red dashed line indicate the V-shape that we discuss in the main text. The bottom row shows face-on surface density projections of stars in the bar in three different age bins as indicated in the top right corner of each panel. We see that younger ages cluster along the bar major axis compared to older ages which have a rounder shape.}
	\label{fig:SFH_auriga1}
\end{figure}

As discussed in Sect. \ref{sect:SFH}, the SFHs along the cuts perpendicular to the major axis of the bar, shown in Fig. \ref{fig:SFH_NGC4981_mass}, have a distinctive V-shape, when examining age versus distance perpendicular to the major axis of the bar. To better understand the origin of this V-shaped age distribution, we explore the SFHs in bars in the Auriga magneto-hydrodynamical cosmological zoom-in simulations \citep{Grand2017}. These are simulations of isolated Milky Way mass halos ($10^{12}$-$2 \times 10^{12}\rm M_{\odot}$) which run from redshift $z$ = 127 to $z$ = 0, with a comprehensive galaxy formation model (see \citealt{Grand2017} and references therein for more details on the simulations). These simulations form disc-dominated galaxies with a significant fraction of 2/3 at redshift $z=0$ having prominent long-lived bars, with properties similar to those of barred galaxies in the local Universe \citep[see][]{BlazquezCalero2019,Fragkoudi2019b}.

In the top left panel of Fig. \ref{fig:SFH_auriga1} we show a face-on mass-weighted age map of one Auriga galaxy (Au18), where we clearly see a prominent bar from the surface density contours. We trace pseudo-slits perpendicular to the major axis of the bar in three different locations, as we did for the TIMER observations, and calculate the mean age of stars along the slits. These are shown in the top right panel of the figure, where we see that within the bar region (horizontal solid lines) there is a dip towards younger ages along the bar major axis. In the second row of the Figure we plot the SFH in each slit, with the leftmost panel corresponding to the black slit and the rightmost panel corresponding to the light grey slit (outer part of the bar). We see that inside the white solid lines, which outline the edge of the bar, there is a characteristic V-shape as the one seen in our observations. Therefore the simulations show a similar V-shaped age distribution inside the bar region as the observations do.\footnote{Au18 experiences a significant merger around $t_{\rm lookback}=8.5\,Gyr$ \citep[see][]{Fragkoudi2019b, Grand2020} which causes a burst of star formation, which can be clearly seen in its SFH as a horizontal line at $\sim$8.5\,Gyr in Fig. \ref{fig:SFH_auriga1}. However, this feature is not relevant for this study here as each galaxy will have its own merger history. Instead, here we focus on the V-shaped SFH in the slits.}

To understand the origin of the V-shape, in the bottom panel of Fig. \ref{fig:SFH_auriga1}, we show the face-on surface density projections of stars in the model in three different age bins: for stars younger than 4\,Gyr (left), stars with ages between 4 and 8\,Gyr (middle) and stars older than 8\,Gyr (right panel). We see that the youngest population has an elongated bar shape, much more so than the oldest population which is rounder. This difference in the shape of the bar according to the age and kinematics of the underlying population was shown using idealised simulations in \citet{Fragkoudi2017} and \citet{Athanassoula2017}, and was termed \emph{kinematic fractionation} by \citet{Debattista2017}. Therefore we see that the younger populations are more clustered along the bar major axis than the oldest populations, due to kinematic fractionation, giving rise to the V-shape we see in the observations.

\subsection{Stellar population properties within the bar radius in cosmological simulations}
\label{sect:comp_Auriga}

In this subsection, we expand our comparison of barred galaxies between TIMER observations and Auriga simulations. We now focus on the spatially resolved 2D stellar population properties that we presented in Sect. \ref{sect:maps} and compare them to the simulations shown in figure 3 of \citet{Fragkoudi2019b}. The latter shows face-on views of ages, metallicities and [$\upalpha$/Fe]-abundances of the inner region ($< 5\,\mathrm{kpc}$) of 5 simulated galaxies.

The authors find that the stellar populations along the bar major axis and at the ends of the bar tend to be younger than populations offset from the bar major axis, which leads to the V-shape in the SFH diagrams that we already discussed. In our 2D light-weighted age maps (our Fig. \ref{fig:age_lw}), this effect is observed in that younger stellar populations are seen at the ends of the bar, although this is not so conspicuous in the mass-weighted maps (Fig. \ref{fig:age_z_mw}), likely because these maps include the additional uncertainties from the light-to-mass conversion and because small differences in very old ages are difficult to measure. On the other hand, our light-weighted age maps do not show clearly older populations on the edges of the bar as these maps are dominated by very young populations from recent and ongoing star formation outside the bar. In this case, the mass-weighted maps, since they highlight older stellar populations better, show at least in IC 1438 and NGC 7755 older populations on the edges of the bar (although, admittedly, this effect is not as clear in the other galaxies). However, as discussed above, the effect is observed in the SFH diagrams as well as along the averaged 1D cuts perpendicular to the major axis.

Furthermore, in the simulations the authors see that bars are more metal-rich than the surrounding discs. This is in good agreement with what we find in the maps of light-weighted metallicities (Fig. \ref{fig:z_lw}). Finally, we find that $\upalpha$-abundances in the simulations also agree very well with our results. Stars of low-$\upalpha$ mainly cluster along the bar major axis, while the surrounding disc shows higher $\upalpha$-enhancement.

\subsection{V-shaped age distribution: where does the time of bar formation fit in?}

The excellent physical spatial resolution of the TIMER data allowed us for the first time to provide observational evidence for a separation of stellar populations by the bar, as it was recently predicted from simulations. In this concept, initially co-planar cold-and-young and hot-and-old stellar populations are mapped into bar-like orbits according to their velocity dispersion, with colder populations getting trapped on more elongated orbits, as opposed to hotter populations which get trapped on rounder orbits.

It is still an open question if and how stars that form \emph{after} the bar, get separated. The key for the morphological separation is the kinematics of the stellar populations or the gas out of which they form, since the bar doesn't have a different gravitational pull on stars just because they are young or old. One possibility is that gas settles into dynamically colder configurations over time and thus stars will form in more elongated orbits. An interesting question is whether there is a second mechanism in which a star that forms in cold orbits would heat, for example through interactions, and therefore migrate to higher energy orbits, i.e. to rounder bar orbits that are further away from the bar major axis.

To shed light on these mechanisms, it would be very interesting to determine the time of bar formation for the galaxies in this sample. This is, in fact, one of the main goals of the TIMER project and it is currently work in progress. The result will give us a horizontal line on the SFHs shown in Figs. \ref{fig:SFH_NGC4981_mass} and \ref{fig:SFH_all}. Everything above that line would have been formed before the bar and everything below the line after bar formation. It will be interesting to see how much of the V-shape is on either side and whether the V-shape is continuous before and after the formation of the bar.

\subsection{Stellar population gradients}

Comparing gradients of stellar population properties, such as age and metallicity, along different axes is a great way of analysing and quantifying the distribution of stellar populations. However, when comparing different results, it is very important to be precise about where and how these gradients were derived.

In our work, we measured gradients along four different axes: (1) MA - along the major axis of the bar between the inner break of the profile (to mask contamination from nuclear structures) and the bar radius, (2) MI - along the extension of the minor axis in the main disc between the inner break and the bar radius, (3) L - along a cut perpendicular to the major axis but offset from the minor axis between the major axis and leading edge of the bar, and (4) T - same as 3 but towards the trailing edge of the bar. This was illustrated in a very simplified way in Fig. \ref{fig:sketch2}.

We found that on average the mass-weighted age gradient along MA is slightly negative but shallow and it is negative and steeper along MI. However, the gradients are positive along L and T. Together, they build a picture in which a bar that is younger along the major axis and older towards its edges is embedded in an even younger main disc. The same can be observed in the top-row panels in Fig. \ref{fig:SFH_auriga1} in the simulated barred galaxy. We speculate that the bar-disc contrast is due to continuous star formation in the outer disc while star formation has been mainly quenched within the bar. This picture is supported by the H$\upalpha$ maps that show ongoing star formation mainly in the outer disc and very little in the bar. The gradient within the bar region is explained by younger stars being trapped into progressively more elongated orbits as discussed in the previous subsection.  Furthermore, the [Mg/Fe]-enhancement that we discussed in Sect. \ref{sect:alphafe} also fits well in this explanation. We found that [Mg/Fe] along bars is lower than in the main disc within the bar radius -- a region that is often called the SFD. It is likely to be populated by old stars from the main disc and, partially, old stars on rounder bar orbits. In the disc outside the bar radius, however, due to continuous star formation, we expect to find a lower [Mg/Fe]-enhancement than in the bar, as reported by \citet{Seidel2016}.

With respect to a comparison between major and minor axis or disc, we found mostly slightly shallower age and metallicity gradients along the major axis, in agreement with results from \citet{Sanchez-Blazquez2011}, \citet{Seidel2016} and \citet{Fraser-McKelvie2019}. This result indicates enhanced radial flattening along the bar likely due to radial movements of stars along very elongated orbits close to the bar major axis. It demonstrates the impact of bars in the radial distribution of stellar populations in galaxies. However, the differences that we observe are smaller than previously reported.

\section{Conclusions}
\label{sect:conclusion}

We have conducted a detailed analysis of spatially resolved stellar populations in galaxy bars within the TIMER project. We have combined mean ages and metallicities, SFHs and [Mg/Fe] abundance ratios with H$\upalpha$ measurements as star formation tracer. We have shown 2D maps as well as averages over pseudo-slits along and perpendicular to the bar major axis that helped us to separate stellar populations across the width of the bars. We have further compared our observational results with cosmological zoom-in simulations from the Auriga project. Our main results can be summarised as follows:

\begin{itemize}

\item Diagrams of SFHs perpendicular to the bar major axes in the MUSE TIMER observations show noticeable 'V-shapes' in the intermediate to old population ($>2\ \mathrm{Gyr}$) which also manifest themselves in positive gradients in profiles of mass-weighted mean ages from the major axis outward. The same shapes are found in the barred galaxies from the cosmological zoom-in simulations of the Auriga project.

These are likely the result of younger and kinematically colder stars being trapped on more elongated orbits -- at the onset of the bar instability -- thus forming a thinner component of the bar seen face-on, and older and kinematically hotter stars forming a thicker and rounder component of the bar. The shapes can also be due to star formation after bar formation, where young stars will form on elongated orbits in the bar region.

\item We showed the imprints of typical star formation processes in barred galaxies on the young age distribution ($<2\ \mathrm{Gyr}$) in the stellar populations. Light-weighted mean stellar ages decrease from the major axis towards the edges of the bar with a stronger decrease towards the leading side. This behavior is especially observed for galaxies that show traces of H$\upalpha$ on the edges of the bar. A stronger effect on the leading side is in accordance with stronger star formation in that region. Furthermore, none of the galaxies in our sample shows significant H$\upalpha$ in the bar except for the presence in central components such as nuclear discs or nuclear rings, at the ends or at the edges of the bar. This result is explained by recent and ongoing star formation in the main disc and in small amounts at the edges of the bar, but not in the region of the bar close to the major axis, probably caused by shear.

\item We found stellar populations in the bars to be in general more metal-rich than in the discs when light-weighted, however, there are notable exceptions, e.g. NGC 4371 and NGC 4984. Except for a prominent peak in the very centre, mass-weighted gradients of mean [Z/H] in the bar are mostly positive but very shallow along the major axis and across the width of the bar. They are on average slightly negative along the extension of the minor axis in the disc. The gradients become more negative, but still shallow, for light-weighted means.

\item Mass-weighted age gradients are negative along both main axes of the bar, but they are shallower along the major axis likely due to orbital mixing in the bar. In general, stellar populations in bars are older than in the discs.

\item Bars are less [Mg/Fe]-enhanced than the surrounding disc. The region of the disc that we probe is mostly within the radius of the bar, which is often called `star formation desert'. We find that [Mg/Fe] is larger in bars than in the inner secularly-built structures but lower than in the SFD. This is indication for a more prolonged or continuous formation of stars that shape the bar structure as compared to shorter formation episodes in the surrounding SFD.

\end{itemize}

\begin{acknowledgements}

We thank Vincenzo Fiorenzo for carefully reading the manuscript and providing a constructive referee report that helped to improve the paper. Based on observations collected at the European Organisation for Astronomical Research in the Southern Hemisphere under ESO programmes 097.B-0640(A) and 060.A-9313(A). JMA acknowledges support from the Spanish Ministerio de Economia y Competitividad (MINECO) by the grant AYA2017-83204-P. J.F-B, AdLC and PSB acknowledge support through the RAVET project by the grant AYA2016-77237-C2-1-P and AYA2016-77237-C3-1-P from the Spanish Ministry of Science, Innovation and Universities (MCIU) and through the IAC project TRACES which is partially supported through the state budget and the regional budget of the Consejer\'\i a de Econom\'\i a, Industria, Comercio y Conocimiento of the Canary Islands Autonomous Community. FAG acknowledges financial support from CONICYT through the project FONDECYT Regular Nr. 1181264, and funding from the Max Planck Society through a Partner Group grant.

\end{acknowledgements}



\bibliographystyle{aa}
\bibliography{NewDatabase}



\begin{appendix}

\section{Alternative comparison of stellar ages, metallicities and [Mg/Fe] abundances}
\label{apx:alt_comp}

\begin{figure*}
	\centering
	\includegraphics[width=0.7\textwidth]{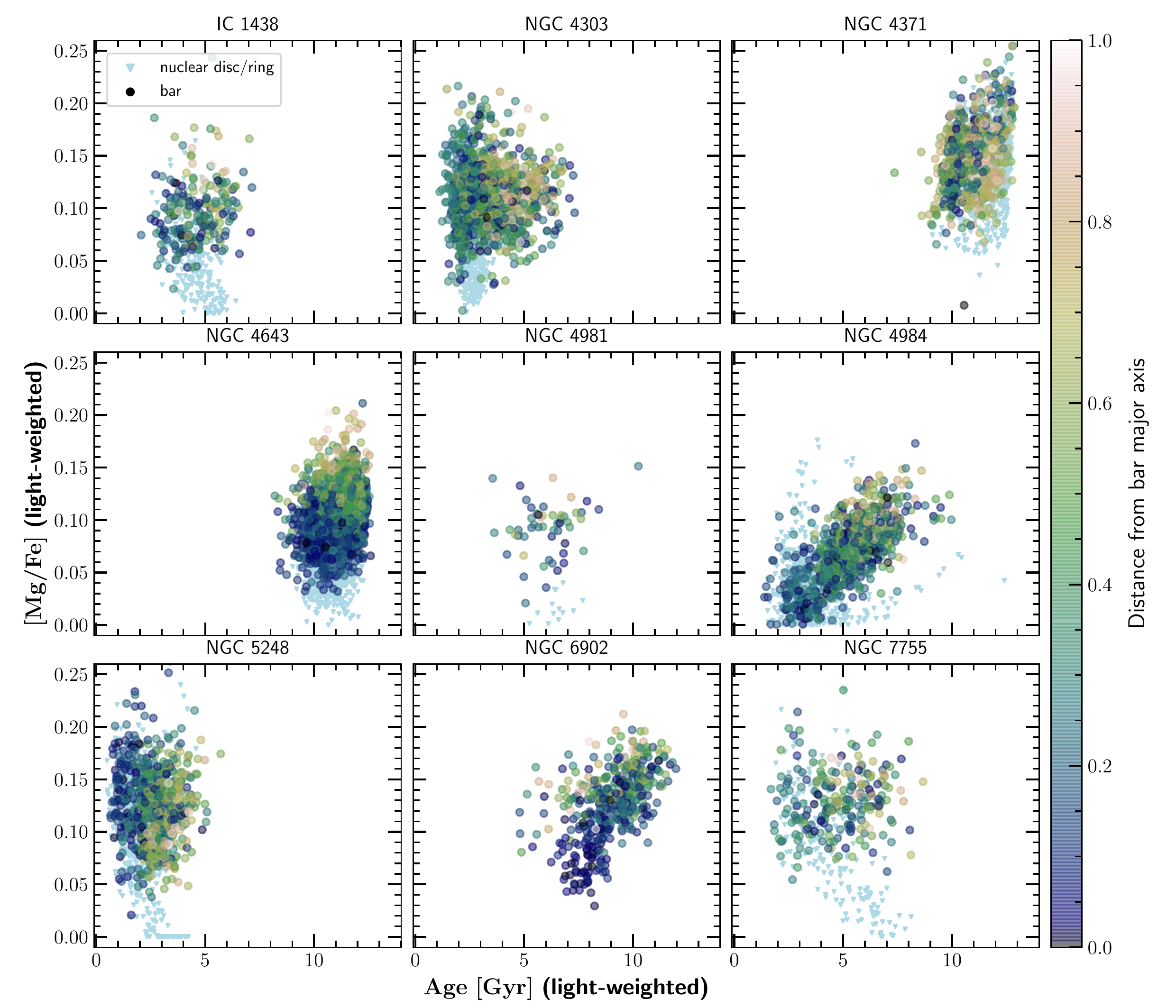}
	\includegraphics[width=0.7\textwidth]{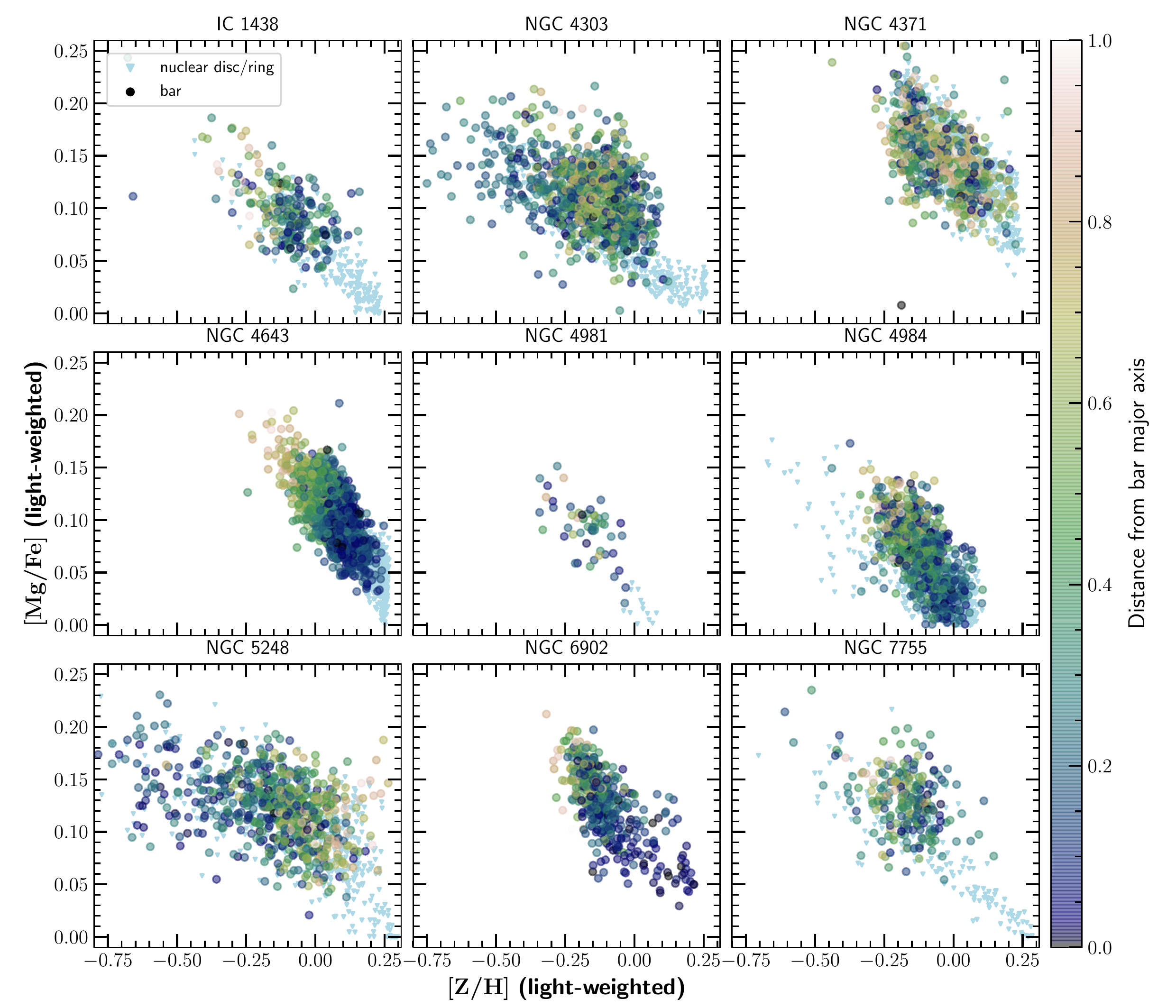}
	\caption{Same as Fig. \ref{fig:ref_NGC4643} but for the complete sample.}
	\label{fig:ref_age_z_lw}
\end{figure*}

\begin{figure*}
	\centering
	\includegraphics[width=0.7\textwidth]{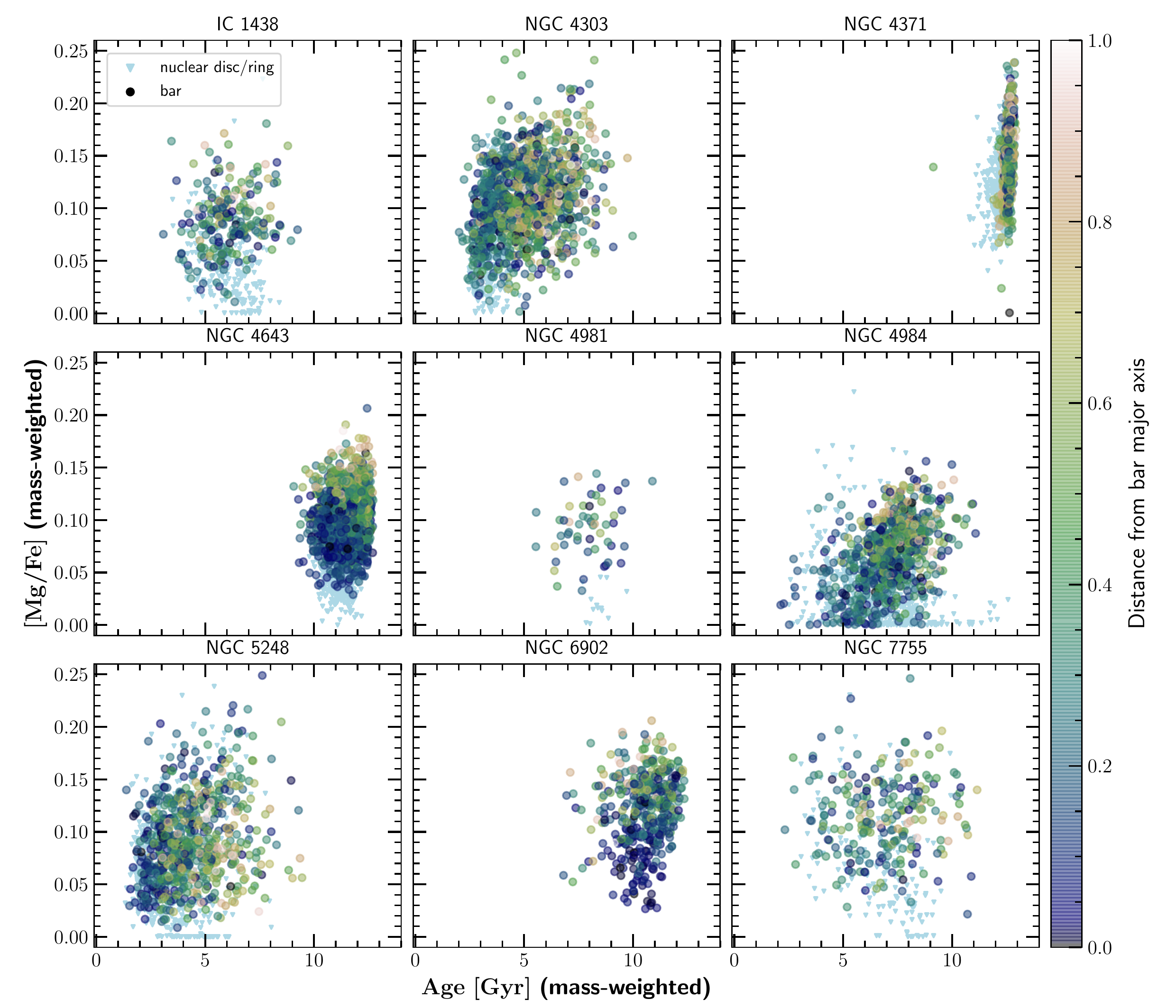}
	\includegraphics[width=0.7\textwidth]{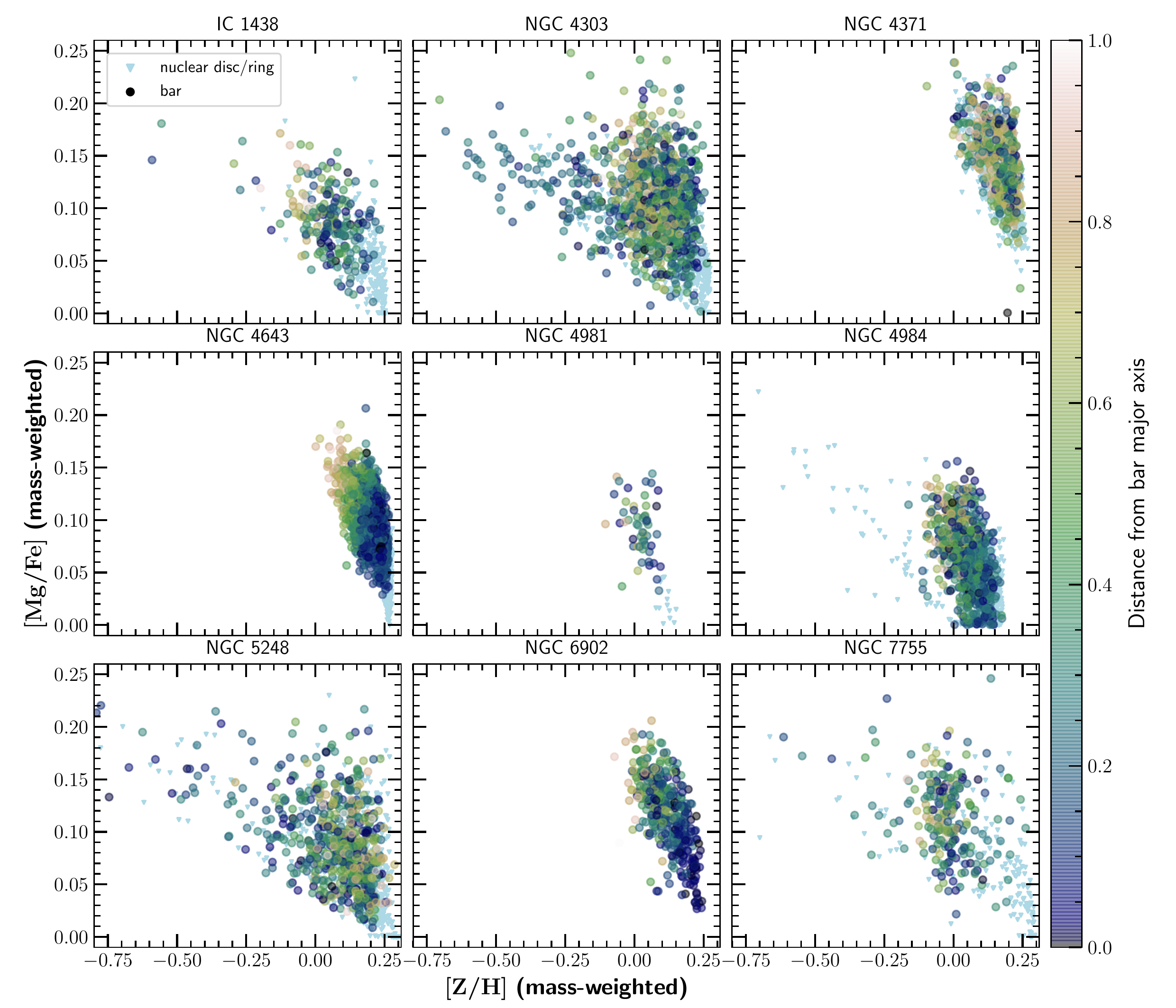}
	\caption{Same as Fig. \ref{fig:ref_age_z_lw} but mass-weighted.}
	\label{fig:ref_age_z_mw}
\end{figure*}

In Sect. \ref{sect:maps}, we presented 2D spatially resolved maps of stellar population properties in form of Voronoi-binned maps and we analysed trends along 1D pseudo-slits in Sect. \ref{sect:4cuts}. These results can also been shown in a different format which we want to employ in this appendix in Figs. \ref{fig:ref_age_z_lw} and \ref{fig:ref_age_z_mw}. An example for NGC 4643 was presented in Fig. \ref{fig:ref_NGC4643} and general trends in the sample discussed in Sect. \ref{sect:alternative_vis}.

\section{Details on mean ages and metallicities along the bar major and minor axis}
\label{apx:agemet_majorminor}

As discussed in Sect. \ref{sect:majorminor}, we derived mean age and metallicity profiles along the bar major and minor axis for all galaxies in the sample. Here, we show one example in Fig. \ref{fig:majorminor_profile} and discuss a few more details of the procedure.

\begin{figure}
	\resizebox{\hsize}{!}{\includegraphics{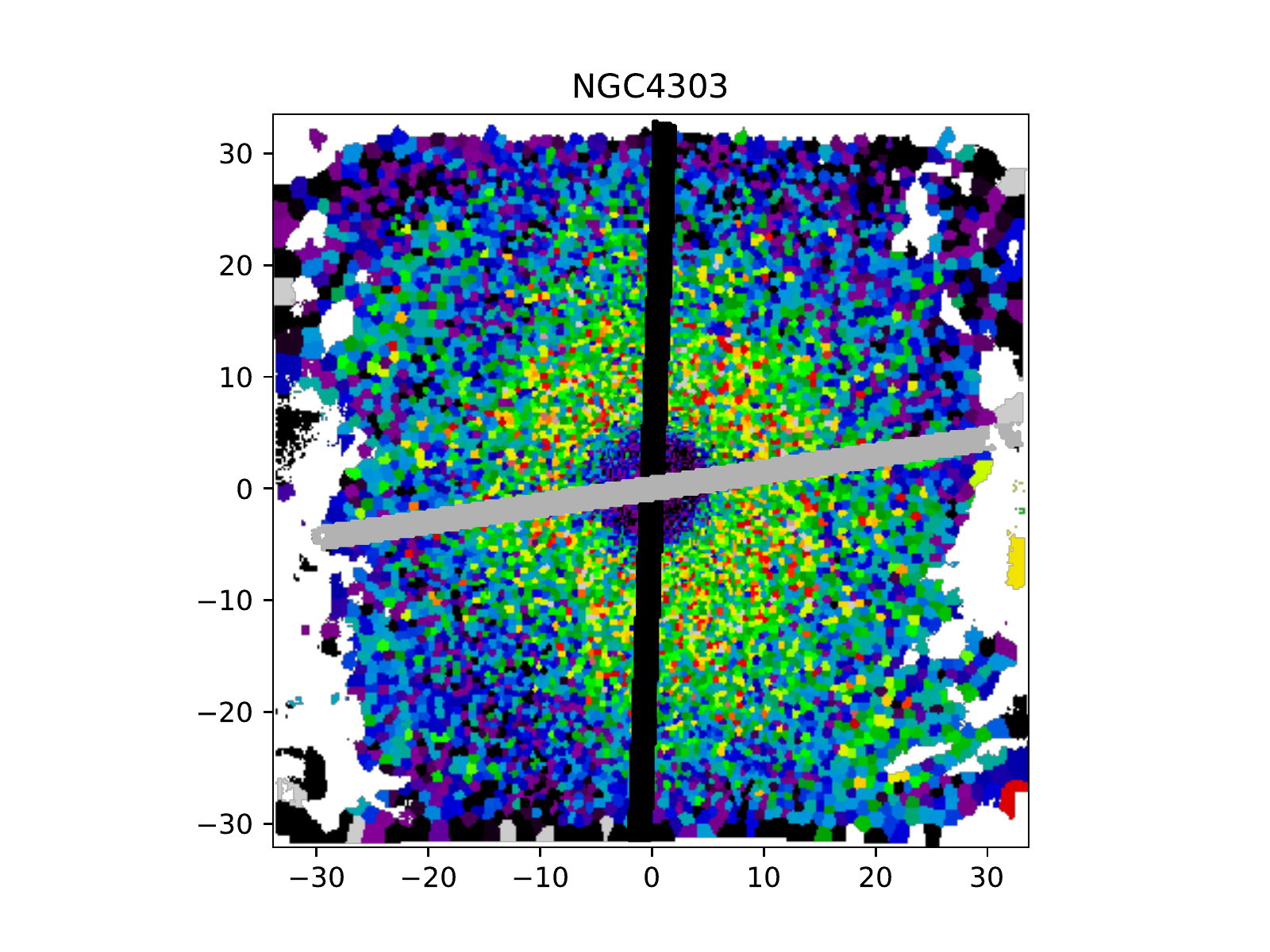}}
	\caption{Sketch to visualise the procedure of extracting age and metallicity profiles along the deprojected bar major and minor axis. Shown is here a map of light-weighted Voronoi-binned mean ages from NGC 4303. Overplotted are pseudo-slits that were used to extract the profile of the major axis (black stripe) and minor axis (grey stripe).}
	\label{fig:majorminor_sketch}
\end{figure}

\begin{figure}
	\resizebox{\hsize}{!}{\includegraphics{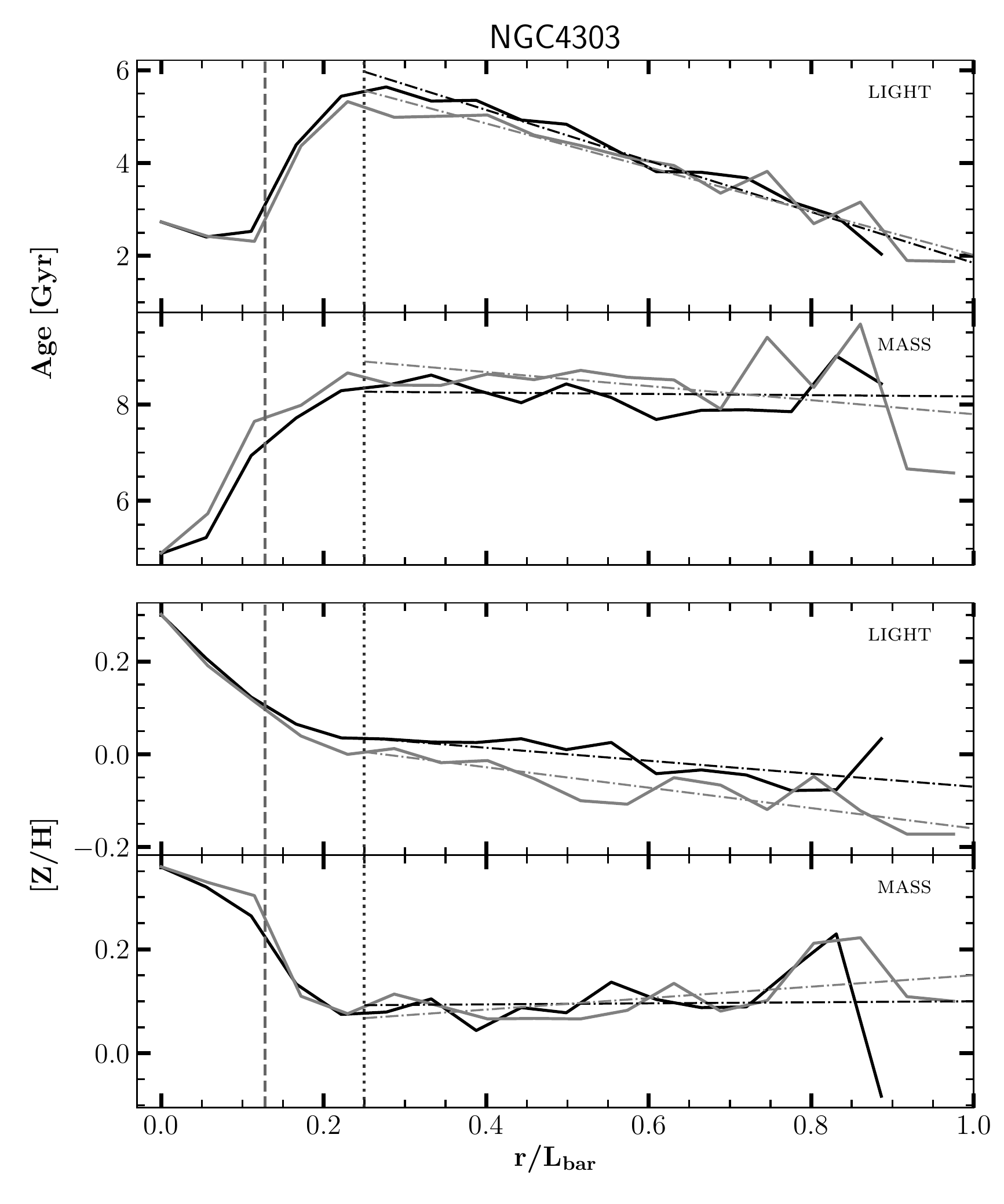}}
	\caption{Mass- and light-weighted mean age and metallicity profiles along the bar major (black) and minor (grey) axis, extended into the disc and normalised by the bar length. This is an example for galaxy NGC 4303. The vertical dashed line shows the position of the nuclear lens component \citep{Herrera-Endoqui2015} and coincides with the first break in the profile. The vertical dotted line marks the position of the second break.}
	\label{fig:majorminor_profile}
\end{figure}

Starting points are the 2D Voronoi-binned maps of mean ages and metallicities derived with \textls{\sc steckmap}. On these maps we determined the position of the major and minor axis of the bar, shown in Fig. \ref{fig:majorminor_sketch}. Note that these axes are not exactly perpendicular to each other on the maps, since we required them to be at $90\degr$ on the deprojected galaxy plane. The deprojection scales were derived from the relative PAs of the axes to the PA of the disc and from the inclination. The average profile was then calculated within pseudo-slits of $2\arcsec$ width in bins of $2\arcsec$ distance along the slit.

An example of these profiles is shown in Fig. \ref{fig:majorminor_profile} for NGC 4303. The distance $r$ along each profile to the centre is the deprojected distance in the galaxy plane and it is divided by the deprojected length of the bar. Two clear breaks are noticeable in all four profiles of this galaxy. The inner break coincides with the position of a nuclear lens component \citep{Herrera-Endoqui2015}. Afterwards follows a transition zone leading to a second break. These breaks are observed in all of our galaxies. In order to measure the slope of each profile (the results of which we showed in Fig. \ref{fig:majorminor_grad}) we decided to use the range of the profile between the second break and the length of the bar. Note that the profiles along the minor axis do not stop at the edge of the bar but continue into the disc.

\section{Details on mean ages and metallicities perpendicular to the bar major axis}
\label{apx:agemet_4cuts}

\begin{figure*}
	\centering
	\includegraphics[width=17cm]{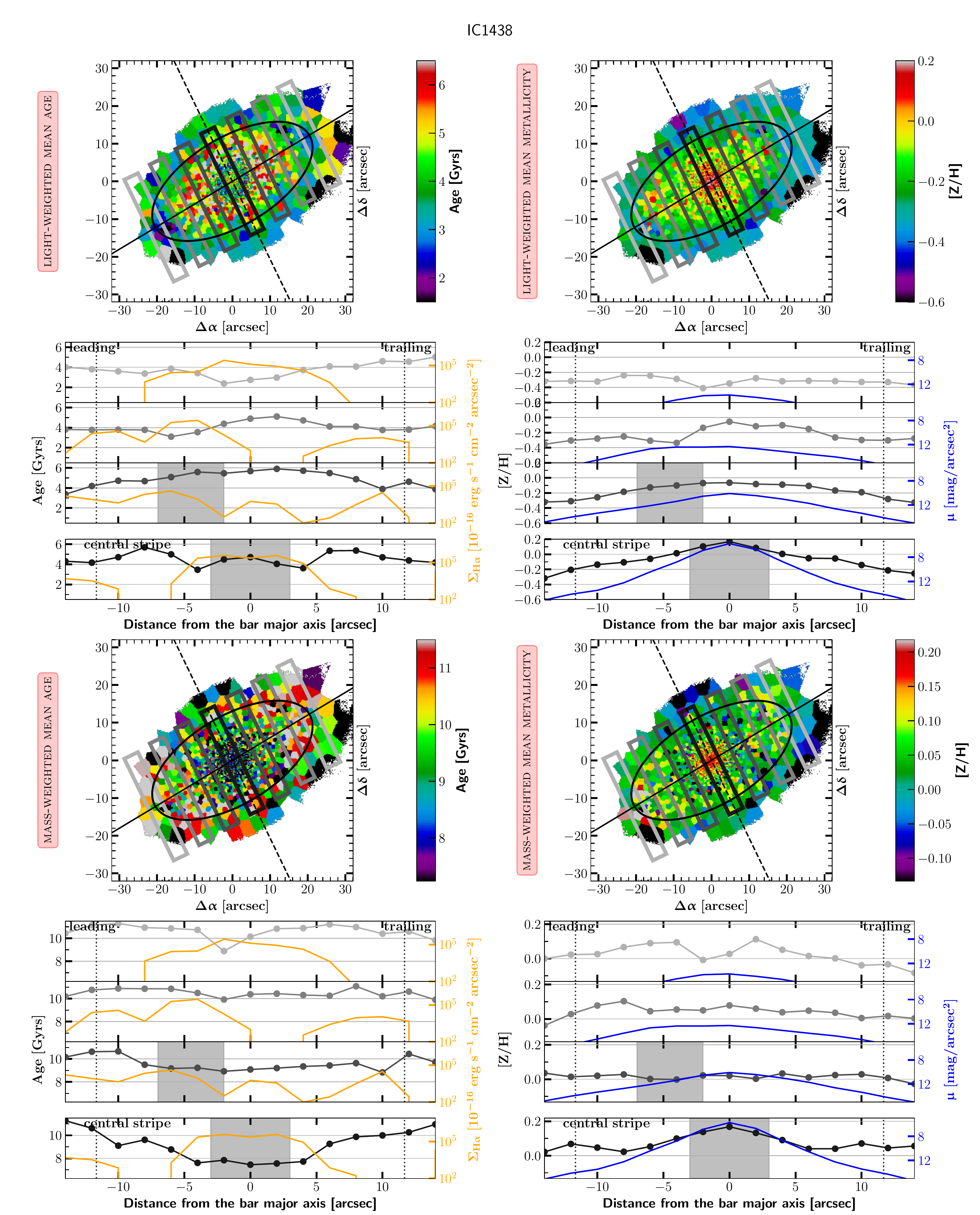}
	\caption{Light- and mass-weighted mean ages and metallicities. The figure splits into four quarters. On the left we show ages and on the right we show [Z/H]. The top shows light-weighted values, the bottom mass-weighted. In each quarter we show a 2D map of Voronoi-binned mean values overplotted with an outline of the bar (black ellipse), the bar major axis (solid line), the minor axis (dashed line) and outlines of the cuts from which we derive the profiles shown below (empty rectangles). Below each map, in four panels, we plot averaged profiles extracted from the corresponding cuts shown in the map. The shades of grey of the profiles correspond to the grey of the rectangles in the map. Also shown are H$\upalpha$ (orange line), total surface brightness (blue line) and the approximate position of dust lines (grey area). Vertical grey dotted lines mark the edge of the bar assuming for simplicity a rectangular shape. Further details can be found in the text.}
	\label{fig:AgeMet_first}
\end{figure*}

In Figs. \ref{fig:AgeMet_first}-\ref{fig:AgeMet_last} we present our detailed analysis of ages and metallicities along four cuts perpendicular to the bar major axis for the complete sample. The main results from the gradients of these profiles were discussed in Sect. \ref{sect:4cuts}.

Equally to the extraction of major and minor axis profiles described in the previous appendix, we start with the 2D maps of light- and mass-weighted mean ages and metallicities as derived from \textls{\sc steckmap}. On these maps we define the bar major and minor axis in projection as previously outlined. Additional to a central cut along the bar minor axis, we define 3 pairs of cuts equally spaced to both sides of the minor axis, such that the last cuts are at the end of the bar. The cuts have a width of $4\arcsec$ and equally spaced bins along the cut every $2\arcsec$. Stripes of the same colour in the figures are averaged in anti-parallel direction. The motivation and a sketch were presented in Sect. \ref{sect:4cuts}.

Additionally, along the cuts, we plot H$\alpha$ densities and total surface brightness. The former is measured along the same cuts from the H$\upalpha$ maps in Fig. \ref{fig:sample_Ha}. The latter is extracted from the total flux within each Voronoi bin during the measurement of the stellar kinematics with \textls{\sc pPXF} (see Sect. \ref{sect:SP_deriv}). For convenience, in order not to overload the figure, we show H$\upalpha$ only in the panels of the left side and the total surface brightness only on the right side, but both can equally be considered for the opposite side as well.

Dust lanes are signatures of cold gas inflows, they are clearly present for most of the galaxies in this sample and can be seen as dark features in the colour maps in Fig. 2 of \citetalias{Gadotti2019}. In our figures, we mark them for reference as grey shaded areas at the approximate position along the profiles.

In Figs. \ref{fig:AgeMet_first}-\ref{fig:AgeMet_last} we present detailed results from this analysis separately for every galaxy in the sample. We do not show individual error bars on the age and metallicity profiles, since an estimation of the uncertainties of the fits with \textls{\sc steckmap} was not performed for all bins within all galaxies. As mentioned in the main text, general uncertainties were studied for a set of 5000 spectra from the TIMER data in Appendix A of \citetalias{Gadotti2019}. Typical values are $0.5$-$1\,$Gyr for age, and $0.005$-$0.010$ for metallicity.

\begin{figure*}
	\centering
	\includegraphics[width=17cm]{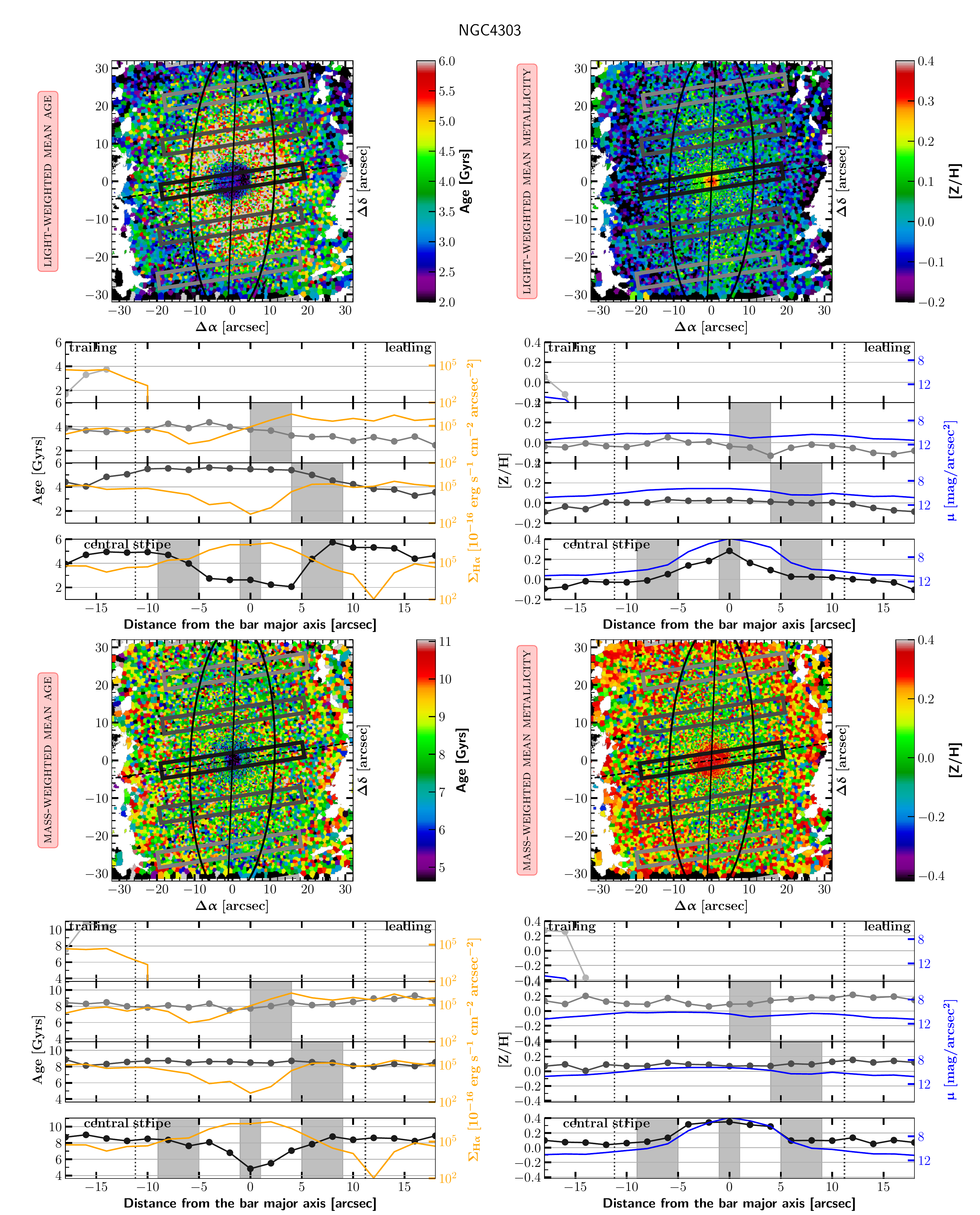}
	\caption{Same as \ref{fig:AgeMet_first}.}
\end{figure*}
\begin{figure*}
	\centering
	\includegraphics[width=17cm]{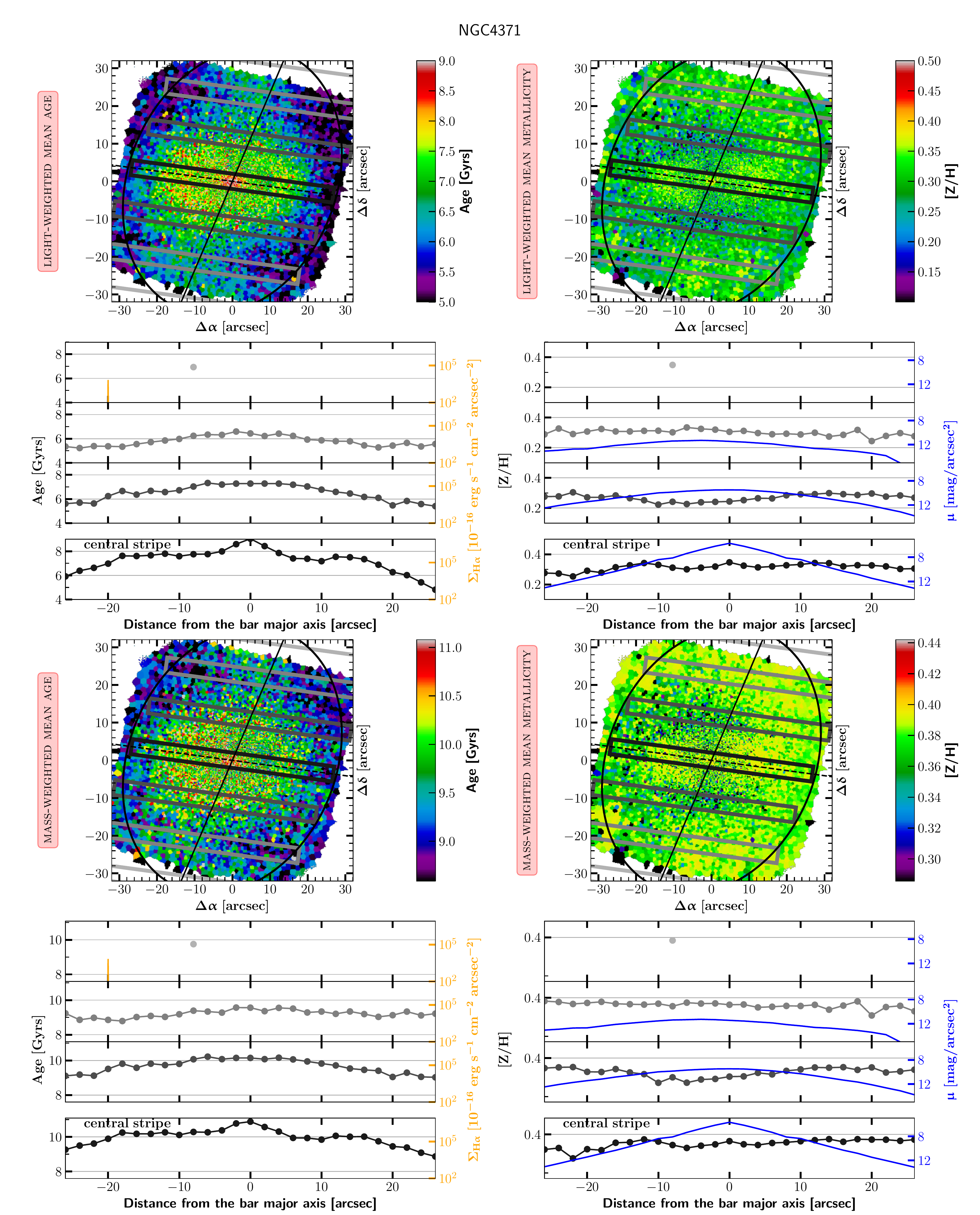}
	\caption{Same as \ref{fig:AgeMet_first}.}
\end{figure*}
\begin{figure*}
	\centering
	\includegraphics[width=17cm]{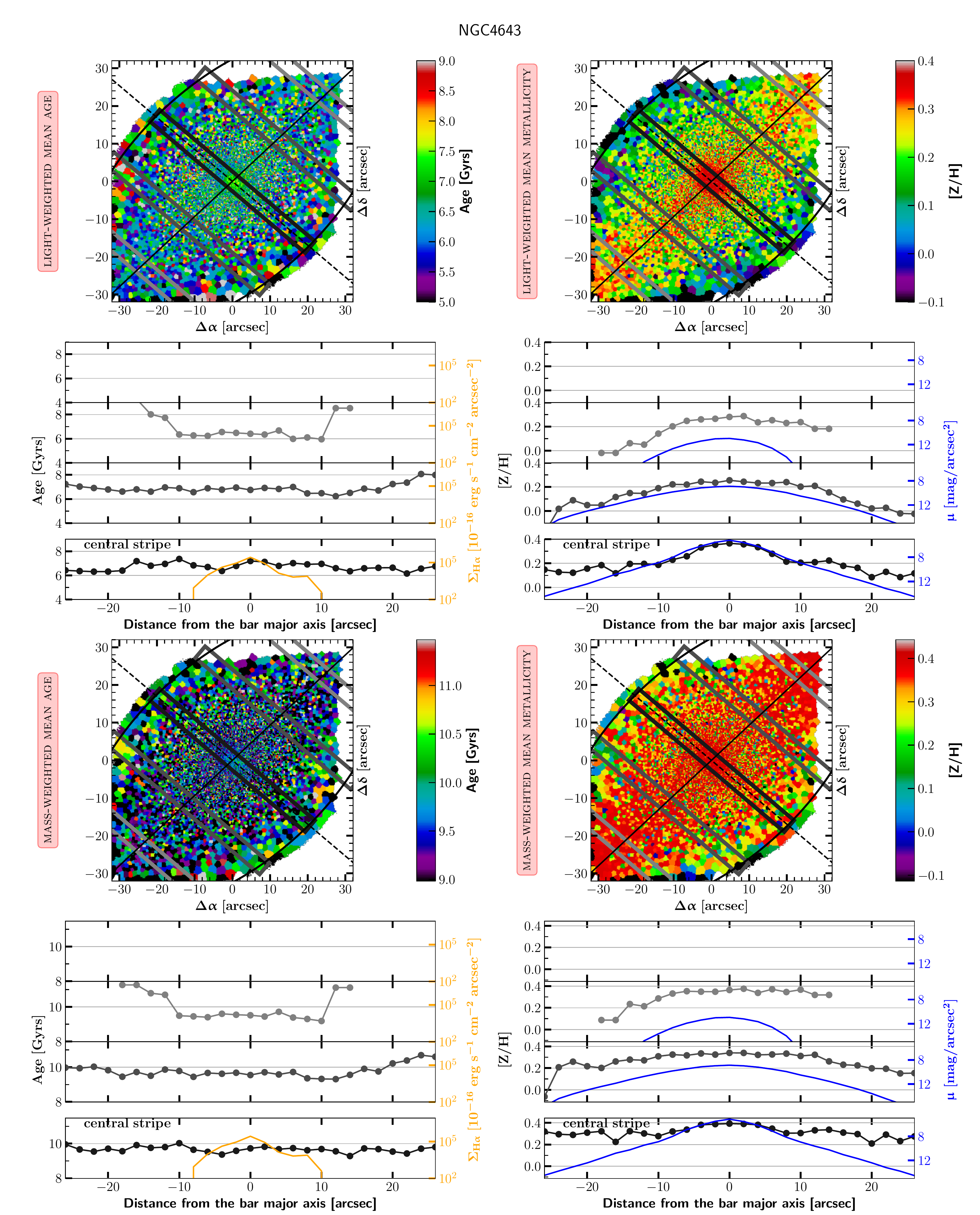}
	\caption{Same as \ref{fig:AgeMet_first}.}
\end{figure*}
\begin{figure*}
	\centering
	\includegraphics[width=17cm]{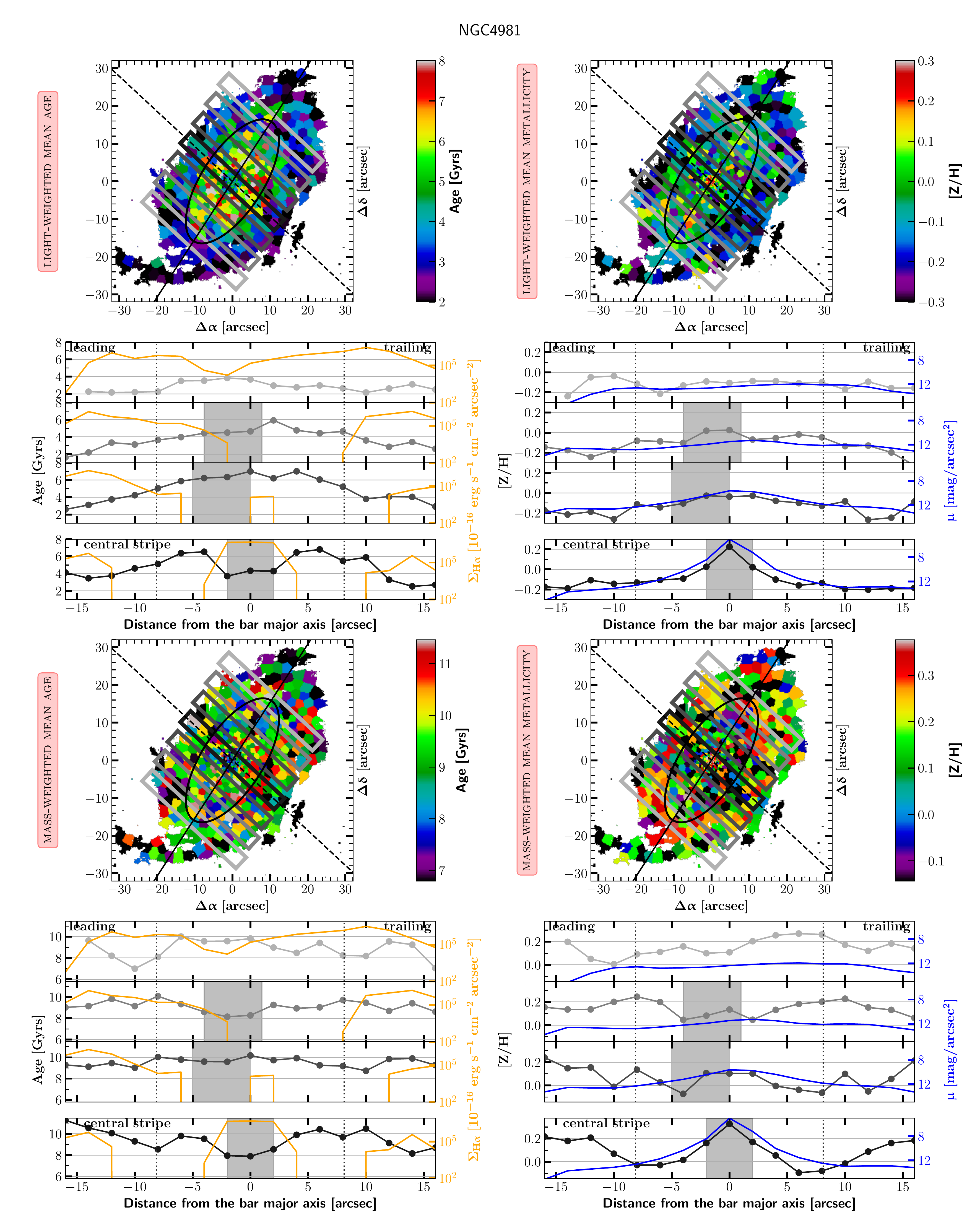}
	\caption{Same as \ref{fig:AgeMet_first}.}
\end{figure*}
\begin{figure*}
	\centering
	\includegraphics[width=17cm]{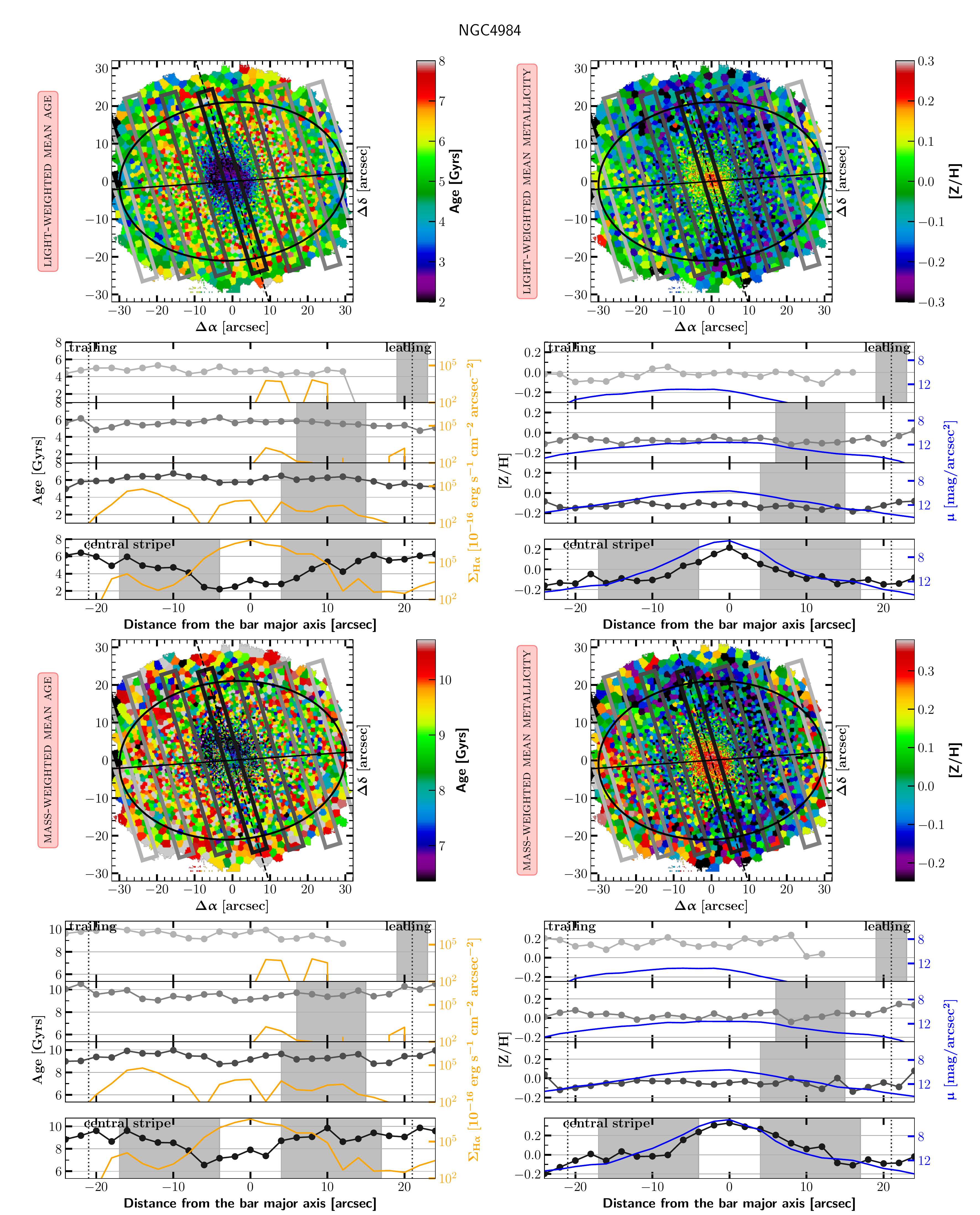}
	\caption{Same as \ref{fig:AgeMet_first}.}
\end{figure*}
\begin{figure*}
	\centering
	\includegraphics[width=17cm]{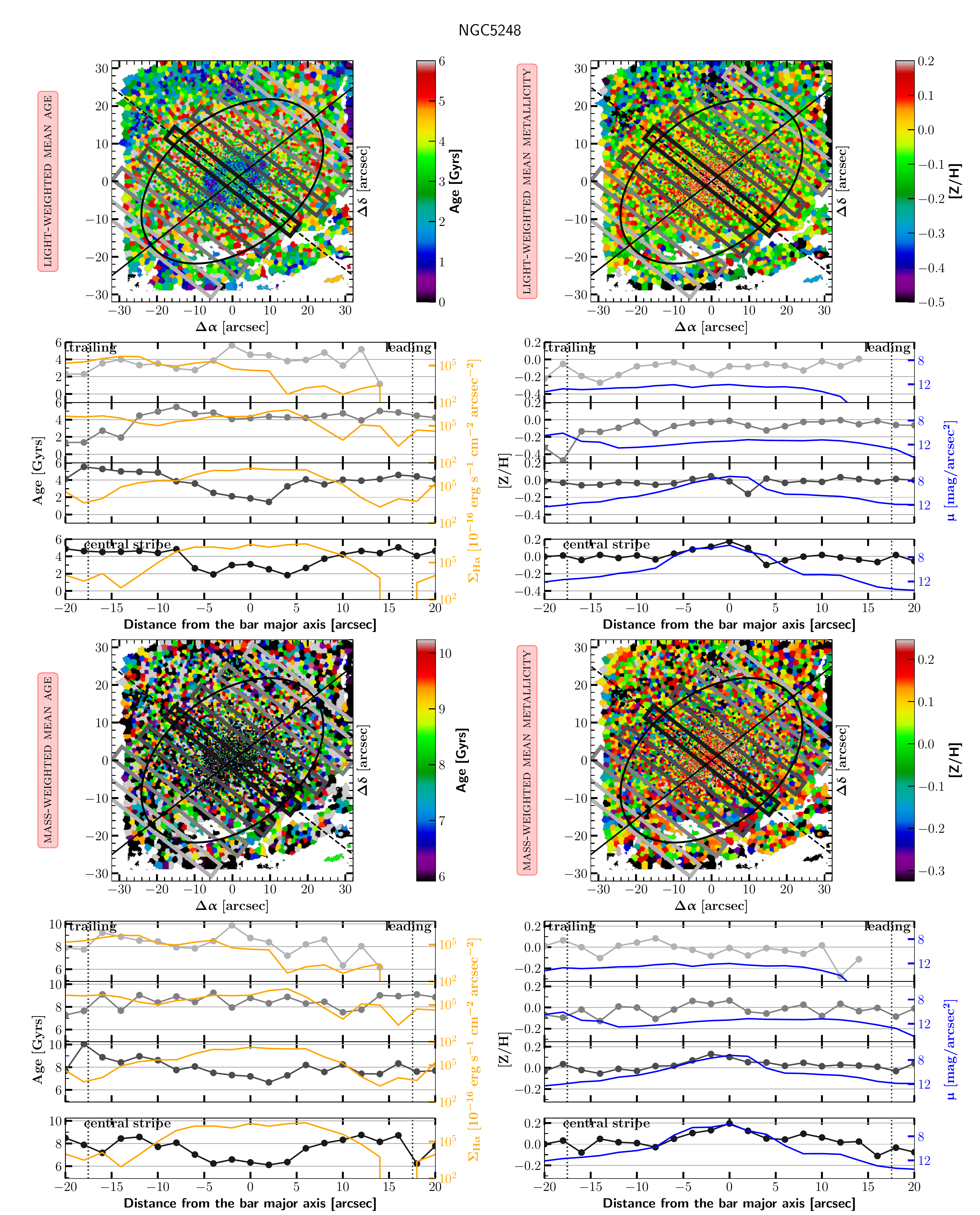}
	\caption{Same as \ref{fig:AgeMet_first}.}
\end{figure*}
\begin{figure*}
	\centering
	\includegraphics[width=17cm]{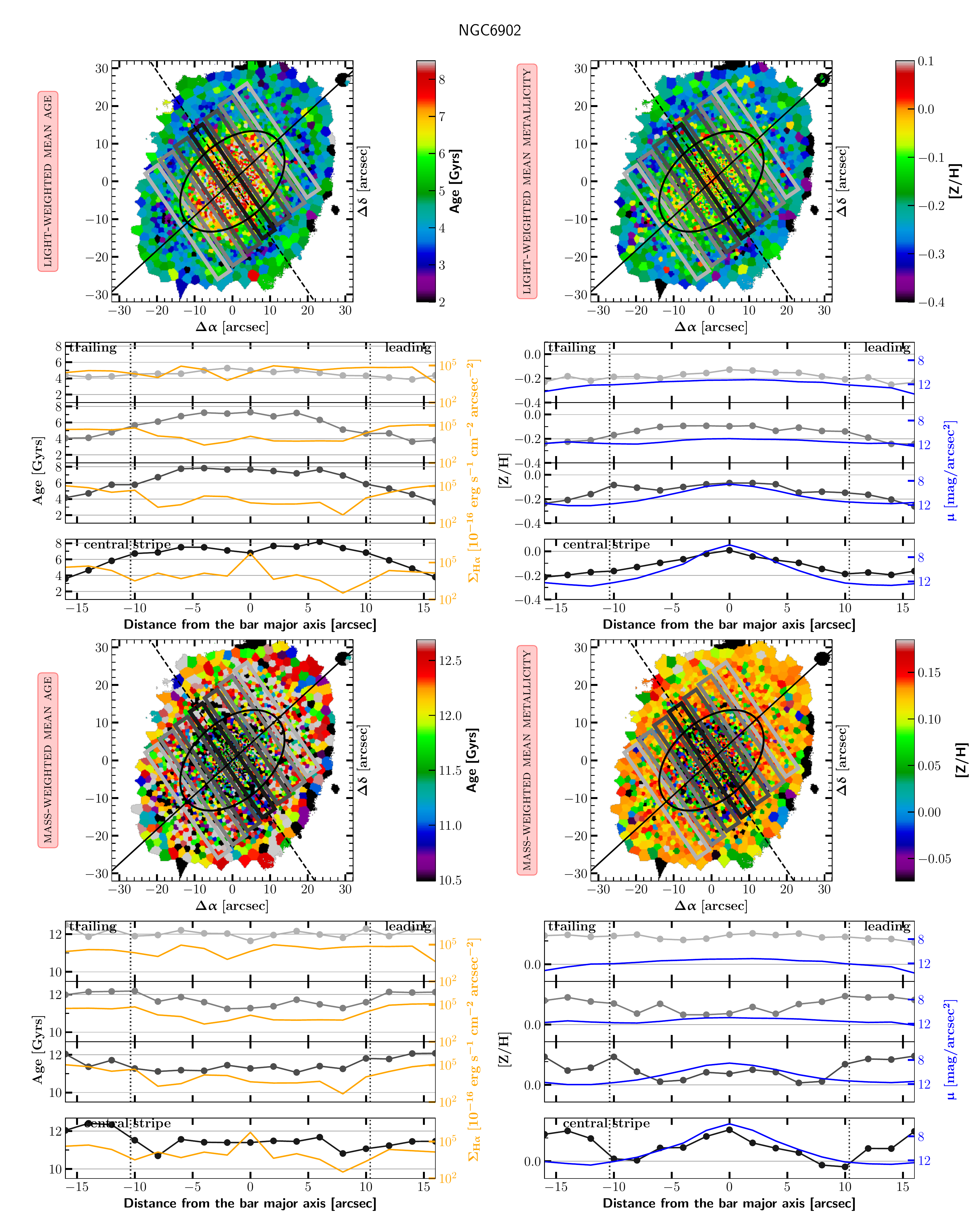}
	\caption{Same as \ref{fig:AgeMet_first}.}
\end{figure*}
\begin{figure*}
	\centering
	\includegraphics[width=17cm]{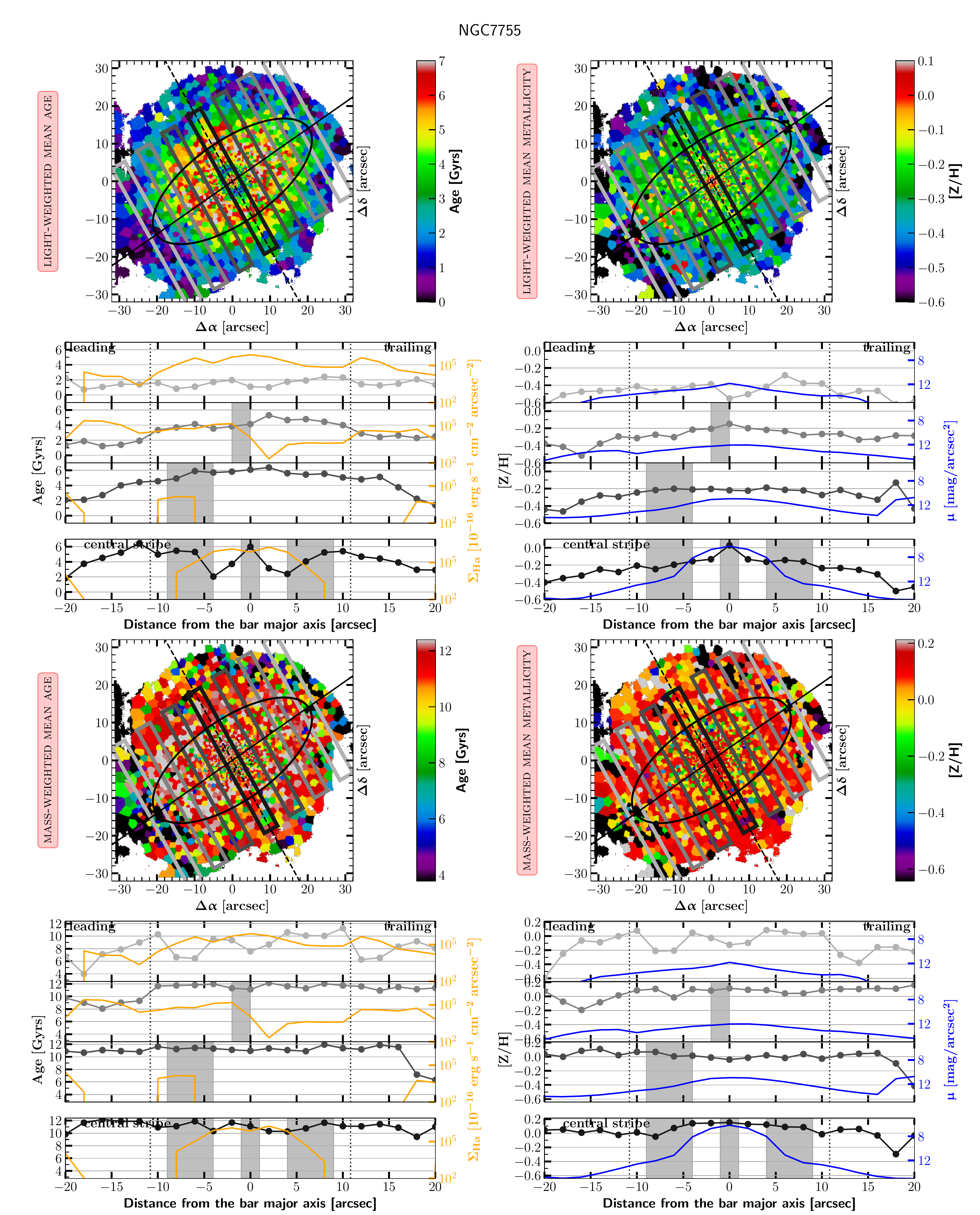}
	\caption{Same as \ref{fig:AgeMet_first}.}
	\label{fig:AgeMet_last}
\end{figure*}

\section{Details on star formation histories}
\label{apx:SFH}

\begin{figure*}
	\centering
	\includegraphics[width=17cm]{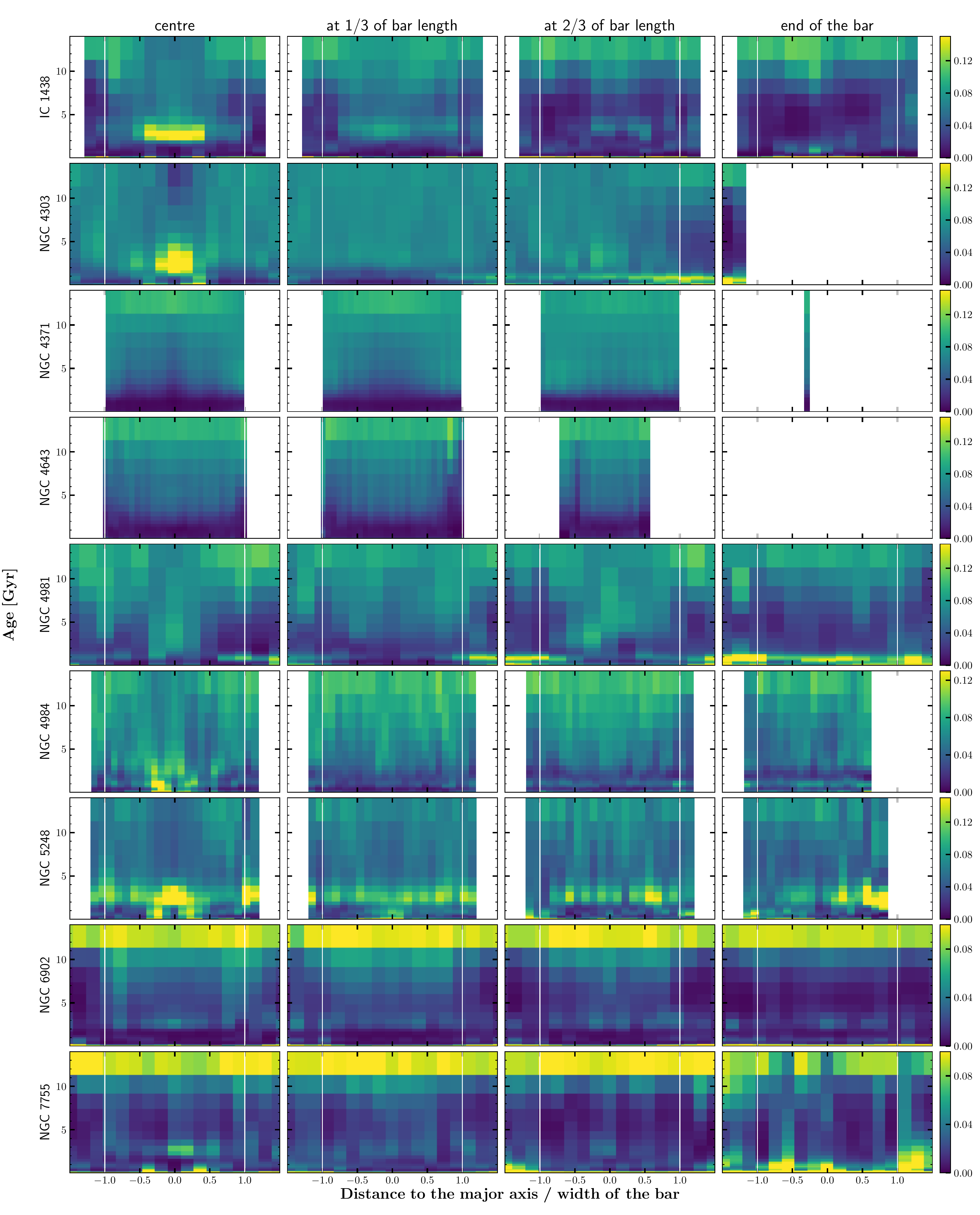}
	\caption{SFHs of the complete sample. Details of the figure are the same as in Fig. \ref{fig:SFH_NGC4981_mass}. The four panels are shown here from left to right in one row for each galaxy. In some cases the colour bar is stretched in order to show fainter details that allow to recognise the `V-shape' discussed in the main text.}
	\label{fig:SFH_all}
\end{figure*}

Star formation histories were derived along the same four cuts perpendicular to the bar major axis for the complete sample. An example was shown for NGC 4981 in Fig. \ref{fig:SFH_NGC4981_mass}, where we also highlighted the apparent `V-shape'. In Fig. \ref{fig:SFH_all}, we show the SFH for all galaxies in the sample. Each row shows one object. The `V-shape' appears in the SFH plots when the edges of the bar are clearly dominated by very old stellar populations while close to the major axis (x=0 in these plots) there is a significant fraction of intermediate-age populations. This shape, sometimes more V-like and sometimes more U-like, can be seen in the galaxies IC 1438, NGC 4643, NGC 4981, NGC 6902, NGC 7755.

\section{Mass-weighted maps of mean ages and metallicities}
\label{apx:agemet_maps}

In this appendix, we present maps of mean stellar ages and metallicities for all galaxies of the sample as in Figs. \ref{fig:age_lw} and \ref{fig:z_lw} but here we show the mass-weighted means.

\begin{figure*}
	\centering
	\includegraphics[width=0.7\textwidth]{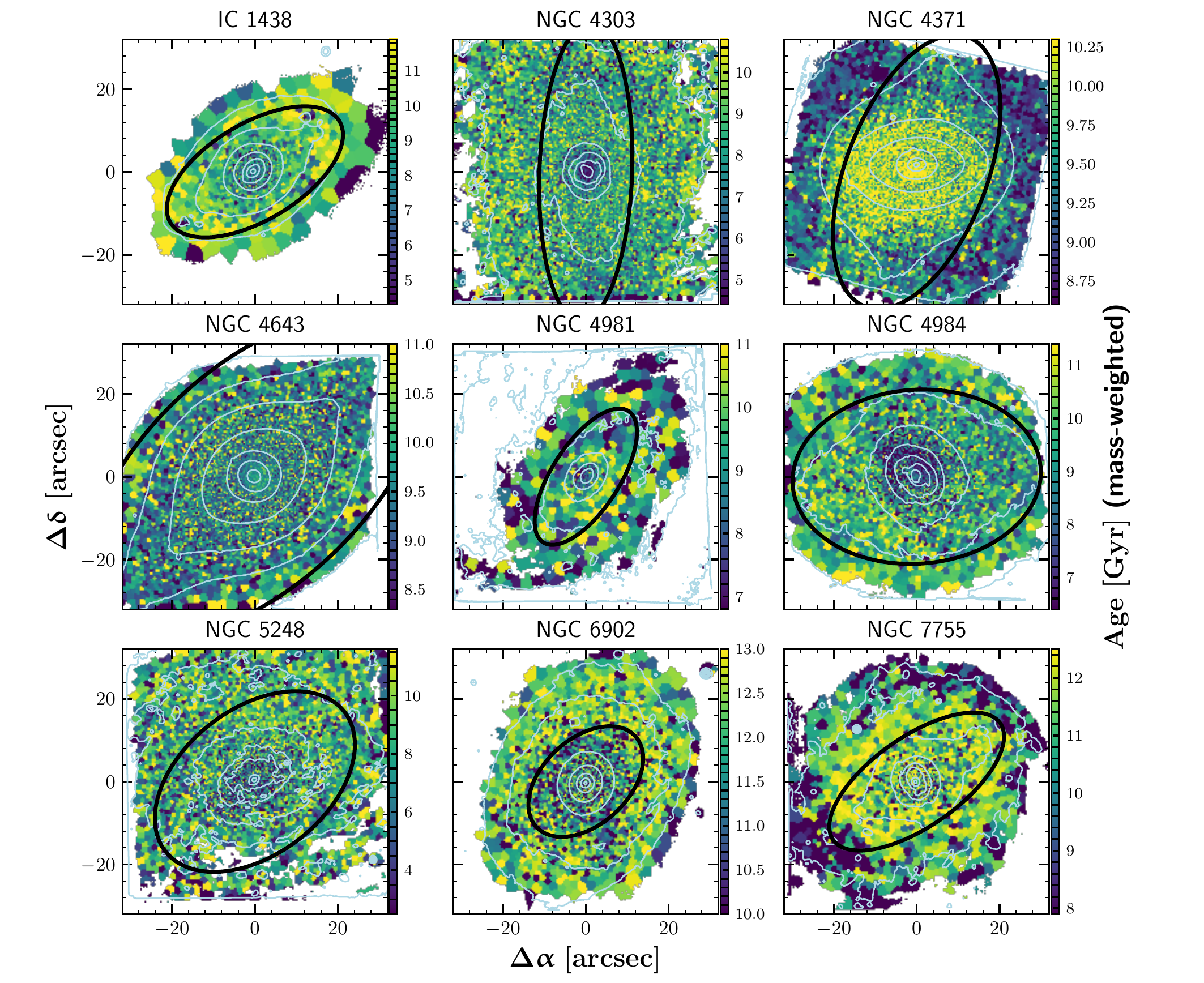}
	\includegraphics[width=0.7\textwidth]{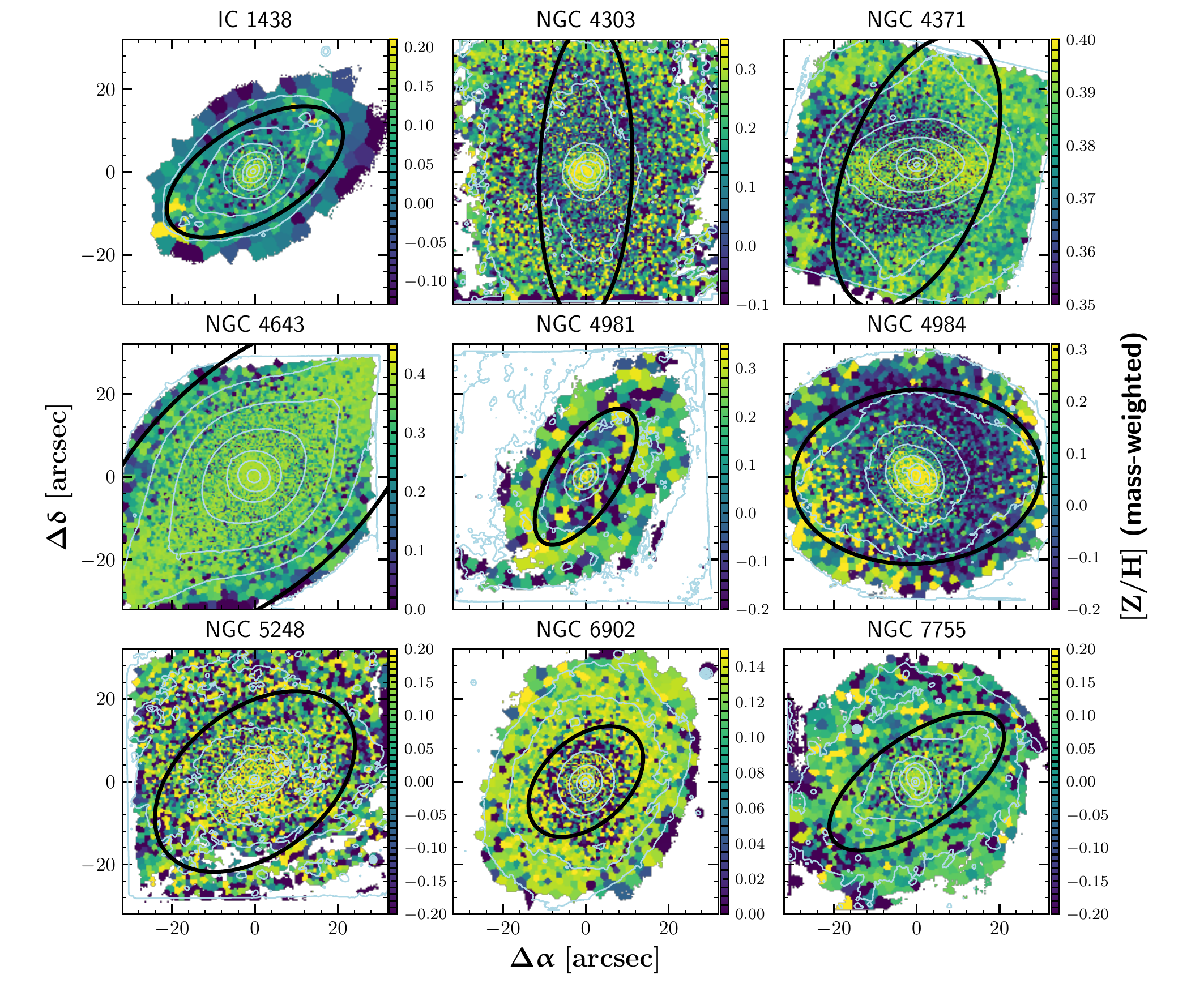}
	\caption{Same as Figs. \ref{fig:age_lw} and \ref{fig:z_lw} but for mass-weighted mean age (top nine panels) and metallicity (bottom nine panels).}
	\label{fig:age_z_mw}
\end{figure*}

\end{appendix}

\end{document}